\documentclass[prb,twocolumn,nofootinbib]{revtex4}
\usepackage{bm}
\usepackage{epsfig}
\newcommand{\etal}[1]{, #1}

\newcommand{\figwidth}{8.6cm}
\newcommand{\figwidthSmall}{6.45cm}
\newcommand{\figheight}{3.5cm}

\renewcommand{\vec}[1]{{\mathbf #1}}
\newcommand{\bsigma}{\bm{\sigma}}
\newcommand{\brho}{\bm{\rho}}
\newcommand{\mubohr}{{\mu_{\rm B}}}
\newcommand{\efermi}{{\varepsilon_{\rm F}}}

\newcommand{\bra}[1]{{\langle #1 |}}
\newcommand{\ket}[1]{{| #1 \rangle}}
\newcommand{\expect}[1]{{\left\langle #1 \right\rangle}}

\newcommand{\spup}{\ket{\!\uparrow}}
\newcommand{\spdown}{\ket{\!\downarrow}}
\newcommand{\spupup}{\ket{\!\uparrow\uparrow}}
\newcommand{\spupdown}{\ket{\!\uparrow\downarrow}}
\newcommand{\spdownup}{\ket{\!\downarrow\uparrow}}

\newcommand{\du}{{\downarrow\uparrow}}
\newcommand{\ud}{{\uparrow\downarrow}}

\newcommand{\downdown}{{\downarrow\downarrow}}

\newcommand{\rhoElem}[1]{\rho_{#1}}
\newcommand{\rhoDotElem}[1]{\dot{\rho}_{#1}}
\newcommand{\rhouu}{\rhoElem{\uparrow}}
\newcommand{\rhodd}{\rhoElem{\downarrow}}
\newcommand{\rhoSS}{\rhoElem{S}}
\newcommand{\rhoSSmax}{\rhoElem{S}^{\rm max}}
\newcommand{\rhouuEq}{\rhoElem{\uparrow}^{\rm eq}}
\newcommand{\rhoddEq}{\rhoElem{\downarrow}^{\rm eq}}
\newcommand{\rhodu}{\rhoElem{\du}}

\newcommand{\rhoSu}{{\rhoElem{S \uparrow}}}

\newcommand{\rhoSd}{{\rhoElem{S \downarrow}}}

\newcommand{\rhoDotuu}{{\rhoDotElem{\uparrow}}}
\newcommand{\rhoDotdd}{{\rhoDotElem{\downarrow}}}
\newcommand{\rhoDotSS}{{\rhoDotElem{S}}}

\newcommand{\rhoDotdu}{{\rhoDotElem{\du}}}

\newcommand{\rhoDotSu}{{\rhoDotElem{S \uparrow}}}

\newcommand{\rhoDotSd}{{\rhoDotElem{S \downarrow}}}
\newcommand{\gammaP}{V}   
\newcommand{\WP}{W}   
\newcommand{\gammadu}{\gammaP_{\du}}
\newcommand{\gammaSu}{\gammaP_{S \uparrow}}
\newcommand{\gammaSd}{\gammaP_{S \downarrow}}
\newcommand{\gammaCTdu}{\gammaP_{\du}^{\rm CT}}

\newcommand{\WSsigma}{{\WP_{S \sigma}}}
\newcommand{\WdS}{{\WP_{\downarrow S}}}
\newcommand{\WSd}{{\WP_{S \downarrow}}}
\newcommand{\SZ}{\bar{S}}
\newcommand{\WdZ}{\WP_{\downarrow \SZ}}
\newcommand{\WZd}{\WP_{\SZ \downarrow}}
\newcommand{\WuS}{\WP_{\uparrow S}}
\newcommand{\WSu}{\WP_{S \uparrow}}
\newcommand{\WuZ}{\WP_{\uparrow \SZ}}
\newcommand{\WZu}{\WP_{\SZ \uparrow}}
\newcommand{\WdSl}[1]{{\WP_{\downarrow S}^{#1}}}
\newcommand{\WSdl}[1]{{\WP_{S \downarrow}^{#1}}}
\newcommand{\WuSl}[1]{{\WP_{\uparrow S}^{#1}}}
\newcommand{\WSul}[1]{{\WP_{S \uparrow}^{#1}}}
\newcommand{\Wud}{\WP_{\ud}}
\newcommand{\Wdu}{\WP_{\du}}
\newcommand{\Win}{\WP_{\rm in}}
\newcommand{\Wout}{\WP_{\rm out}}

\newcommand{\WCTdu}{{{\WP}_{\du}^{\rm CT}}}
\newcommand{\WCTud}{{{\WP}_{\ud}^{\rm CT}}}
\newcommand{\Womegarf}{\WP_{\omegarf}}
\newcommand{\WoMax}{\WP_{\omegarf}^{\rm max}}
\newcommand{\Eu}{E_{\uparrow}}
\newcommand{\Ed}{E_{\downarrow}}
\newcommand{\ES}{E_{S}}
\newcommand{\ETp}{E_{T_+}}
\newcommand{\EZ}{E_{\SZ}}
\newcommand{\ESs}[1]{\Delta_{S#1}}
\newcommand{\ESu}{\Delta_{S\uparrow}}
\newcommand{\ESd}{\Delta_{S\downarrow}}

\newcommand{\EdZ}{\Delta_{\downarrow\SZ}}
\newcommand{\EuZ}{\Delta_{\uparrow\SZ}}
\newcommand{\EStark}{\delta\epsilon}
\newcommand{\gammaTZero}{\gamma}
\newcommand{\gammaTZeroAvg}{\gamma}
\newcommand{\gls}[2]{\gammaTZero^{#2}_{#1}}
\newcommand{\gul}[1]{\gammaTZero^\uparrow_{#1}}
\newcommand{\gdl}[1]{\gammaTZero^\downarrow_{#1}}
\newcommand{\guO}{\gul{1}}
\newcommand{\guT}{\gul{2}}
\newcommand{\gdO}{\gdl{1}}
\newcommand{\gdT}{\gdl{2}}
\newcommand{\gO}{\gammaTZero_1}
\newcommand{\gT}{\gammaTZero_2}
\newcommand{\guAvg}{\gammaTZeroAvg^\uparrow}  
\newcommand{\gdAvg}{\gammaTZeroAvg^\downarrow}  
\newcommand{\gAvg}{\gammaTZeroAvg}  

\newcommand{\fls}[2]{f_{#1}(\Delta_{S #2})}
\newcommand{\ful}[1]{\fls{#1}{\uparrow}}
\newcommand{\fdl}[1]{\fls{#1}{\downarrow}}
\newcommand{\fZls}[2]{f_{#1}(\Delta_{#2\SZ})}
\newcommand{\fZul}[1]{f_{#1}(\EuZ)}
\newcommand{\fZdl}[1]{f_{#1}(\EdZ)}
\newcommand{\tmeas}{t_{\rm meas}}
\newcommand{\hgu}{\hat{\gammaTZero}^{\uparrow}}
\newcommand{\hgd}{\hat{\gammaTZero}^{\downarrow}}

\newcommand{\Dm}{\Delta\mu}

\newcommand{\kmudot}{\mu_{\rm dot}}
\newcommand{\Eadd}{E_{\rm add}}

\newcommand{\Hlead}{{H_{\rm lead}}}
\newcommand{\Hdot}{{H_{\rm dot}}}
\newcommand{\Hrf}{{H_{\rm ESR}}}
\newcommand{\Llead}{{L_{\rm lead}}}
\newcommand{\Ldot}{{L_{\rm dot}}}
\newcommand{\Lrf}{L_{\rm ESR}}
\newcommand{\HrfOsc}{{\cos{(\omegarf  t)}}}
\newcommand{\HrfConst}{{\Delta_x}} 
\newcommand{\omegarf}{{\omega}}
\newcommand{\HrfOffDiag}{{\HrfConst \HrfOsc}}
\newcommand{\HrfRotConst}{\Delta_\perp}  

\newcommand{\HDD}{H_{\rm DD}}
\newcommand{\HDL}{H_{\rm DL_2}}
\newcommand{\tDD}{t_{\rm DD}}
\newcommand{\tDL}{t_{\rm DL_2}}

\newcommand{\Ibg}{\bar{I}_0}

\newcommand{\rhoSys}{\rho_{\rm D}}
\newcommand{\brhoSys}{\brho_{\rm D}}
\newcommand{\rhoSysIA}{\rho_{\rm D}^{\rm I}}
\newcommand{\rhoDotSys}{\dot{\rho}_{\rm D}}
\newcommand{\brhoDotSys}{\dot{\brho}_{\rm D}}
\newcommand{\rhoDotSysIA}{\dot{\rho}_{\rm D}^{\rm I}}

\newcommand{\bathLbl}{{\rm B}}
\newcommand{\TrB}{{{\rm Tr}_\bathLbl \,}}

\newcommand{\rBNot}{{\rho_{\rm R}^0}}
\newcommand{\rSNot}{\rhoSys(0)}
\newcommand{\LV}{{L_T}}

\newcommand{\textQdot}{\setlength{\unitlength}{1pt}}
\newcommand{\captionQdot}{\setlength{\unitlength}{0.9pt}}

\newcommand{\qdot}[1]{\begin{picture}(14,11)
    \put(6,3.6){\circle{13}}
    \put(6,3.6){\makebox(0,0){$#1$}}
    \end{picture}}
\newcommand{\edot}[1]{\begin{picture}(7,9) 
    \put(3,3.6){\makebox(0,0){$#1$}}
    \end{picture}}
\newcommand{\qddot}[2]{\qdot{#1}$_{\!1}$\qdot{#2}$_{\!2}$}

\begin{document}


\title{ 
Single Spin Dynamics and Decoherence in a Quantum Dot via Charge Transport
}

\author{Hans-Andreas \surname{Engel}}
\email{Hans-A.Engel@unibas.ch}

\author{Daniel \surname{Loss}}
\email{Daniel.Loss@unibas.ch}

\affiliation{Department of Physics and Astronomy, University of Basel,
 Klingelbergstrasse 82, CH-4056 Basel, Switzerland}

\begin{abstract}
We investigate the spin dynamics of a  quantum dot with a spin-$\frac{1}{2}$
ground state
in the Coulomb blockade regime and in the presence of a magnetic rf field
leading to electron spin resonances (ESR).
We show that by coupling the dot to leads,
spin properties on the dot can be accessed via the charge current in the
stationary and non-stationary limit.
We present a microscopic derivation of the current and the master equation
of
the dot using superoperators,
including contributions to decoherence and energy shifts
due to the tunnel coupling.
We give a detailed analysis of sequential
and co-tunneling currents,
for linearly and circularly oscillating ESR fields,
applied in cw and pulsed mode.
We show that the sequential tunneling current
exhibits a spin satellite peak
whose linewidth gives a lower bound on
the  decoherence time $T_2$ of the spin-$\frac{1}{2}$ state on the dot.
Similarly, the spin decoherence
can  be accessed also in the cotunneling regime via ESR induced spin flips.
We show that the conductance ratio of the spin satellite peak and the
conventional peak due to sequential tunneling saturates at the universal conductance ratio of 0.71
for strong ESR fields.
We describe a double-dot setup which generates spin dependent tunneling and 
acts as a current pump (at zero bias), and as a spin inverter which
inverts the spin-polarization of the current,
even in a homogeneous magnetic field.
We show that Rabi oscillations of the dot-spin induce
coherent oscillations in the time-dependent current.
These oscillations are observable in
 the time-averaged current as function of ESR pulse-duration,
and they allow one
to access the spin coherence
directly in the time domain.
We analyze the measurement and read-out process of the dot-spin
via currents in spin-polarized leads
and identify measurement time and efficiency
by calculating the counting statistics, noise, and the Fano factor.

\end{abstract}


\maketitle

                       \section{Introduction}
The coherent control and manipulation of the electron spin has become the
focus
of an increasing number of
experiments.${}^{1-8}$
{}From measurements it has become evident  that the phase coherence of
electron spins in 
semiconductors 
can be robust over unusually long times, exceeding 100's of
nanoseconds.\cite{Kikkawa}
Thus, spins of electrons are
suitable candidates
for applications in the field of spintronics, in particular for
quantum information 
 processing.${}^{9-18}$
This has made it desirable to understand in more detail
the coherent behavior of single electron spins which are confined to
nanostructures such as quantum dots, molecules, or atoms, and to point to ways of 
how to access  the coherence time of a single spin experimentally.
It is the goal of this work to address
this issue and to propose and analyze  transport
scenarios involving    a quantum dot attached to leads and with a spin-1/2 ground state.

We first remind ourselves of some basic notions in spin dynamics.
When the electron spin is exposed to an external magnetic field, this leads
 to a Zeeman splitting, and
the spin dynamics is described by the standard Bloch equations.\cite{Abragam}
These are characterized by two time scales: the (longitudinal) relaxation
time $T_1$ and the decoherence time $T_2$ (transverse relaxation).
The spin relaxation time $T_1$ describes the
lifetime of an excited spin state, aligned along the external field,
and is classical in the sense that its definition
does not involve the concept of quantum superpositions.
Such a $T_1$ time of a spin in
a single quantum dot
was measured recently via transport and was shown to be longer than a few
microseconds,\cite{Fujisawa}
in agreement with calculations.\cite{KhaetskiiNazarov}
On the other hand,
the spin decoherence time $T_2$ gives the time  over which
a superposition of opposite spin states of a single electron
remains coherent. 
Thus, coherent manipulations of electron spins,
e.g., gate operations for quantum computation,
must be performed faster than $T_2$.
We note that quite generally $T_2 \leq T_1$.\cite{Abragam}
Thus, from the sole knowledge of $T_1$
no lower bound for $T_2$ follows.
It is thus of fundamental interest to investigate possibilities
of how to gain access to the decoherence time $T_2$ for  a single spin
confined to a quantum dot.

The loss of phase coherence of many but independent spins is described by the
dephasing time\cite{Kikkawa} $T_2^{*}$, 
 where inhomogeneities in the Zeeman terms 
 lead to a further suppression of phase coherence for the ensemble 
 but not necessarily
 for an individual spin, thus $T_2^{*} \leq T_2$.
In recent experiments, $T_2^{*}$
was measured in bulk GaAs
by using ultrafast time-resolved optical methods, yielding
 values for $T_2^{*}$ exceeding 100 ns.\cite{Kikkawa}

However,  the measurement of
the  decoherence time $T_2$ for a single spin has---to our knowledge---not been reported yet
(although it is expected to be within experimental reach given the known single-photon
sensitivity).
A first step into this direction are
spin echo measurements on an ensemble of spins,
where dephasing due to inhomogeneities of the  magnetic field is eliminated.
Indeed, such measurements being performed more than thirty years ago
on P donors in Si, reported $T_2$ times up to $500\:{\rm \mu}s$.\cite{T2Si}
However, it appears 
desirable to have a more direct method for single spin measurements.
{}To achieve this via direct coupling to the magnetic moment of the spin
is rather challenging due to the extremely small magnetic moment,
although it is believed to be within reach using cantilever techniques.\cite{Hammel}
Here we concentrate on a further approach based on  transport measurements. The
key idea is to exploit
the Pauli principle which connects spin and charge of the electron
so intimately that all spin properties
can be accessed via  charge and  charge currents, especially
in the Coulomb blockade regime\cite{kouwenhoven} of a quantum dot attached to leads.
Indeed, concrete scenarios based on such a spin-to-charge conversion
have been proposed in the past,\cite{Loss97,Recher,EL,QCReview,GL} and it is our goal here to 
further elaborate on these concepts, and to report on a variety of new results we have obtained.

There are two classes of 
spin decoherence contributions we have to distinguish in the following.
First, rare tunneling events of electrons onto and off the dot  change 
 the spin state of the dot and in this way contribute to the
decoherence of the dot spin. 
We account for this decoherence  microscopically in terms of a
  tunneling Hamiltonian.
Second, there are 
 intrinsic decoherence contributions from  processes
 which persist even if the dot is completely isolated from the leads.
This decoherence is taken
 into account phenomenologically in the master equation developed
in this work, with an intrinsic
decoherence rate
$T_2^{-1}$. 
The goal then is to show that this $T_2$ time can be extracted
via current measurements, regardless of the microscopic processes leading
to $T_2$.
Such a phenomenological approach
to intrinsic decoherence makes  
 the purpose of our considerations clearer and is applicable to different
 types of decoherence mechanisms e.g. based on hyperfine and spin-orbit couplings.
The microscopic study of such intrinsic decoherence, being an important subject in
its own right, is not addressed in the present work.

The outline of this paper is as follows.
In Sec.~\ref{secQdot}, we define the system of interest,
 a quantum dot with spin-1/2 ground state
in the Coulomb blockade regime tunnel-coupled to leads, and in
 the presence of an electron spin resonance (ESR) field.
We derive the (generalized) master equation for the low energy dot states
 in the sequential and co-tunneling regime
 by evaluating the tunnel coupling to the leads microscopically
 in order to obtain tunneling rates, decoherence rates, and
 energy (Stark) shifts. For this we need to include diagonal and off-diagonal
matrix elements of the reduced density operator.
The stationary current through the dot and its dependence on 
 the ESR field is discussed in Sec.~\ref{secStationaryCurrent}.
We find a spin satellite peak in the sequential tunneling current,
 whose linewidth as function of the ESR frequency gives a lower bound 
 for the $T_2$ time. Thus,  via the stationary
current, the $T_2$ time can be accessed in a regime that is experimentally accessible,
as will be demonstrated by concrete numerical examples.
We show that the ratio of this satellite peak and the
 main peak saturates at a universal conductance ratio
 for strong ESR fields.
In Sec.~\ref{secZeroOne}, we extend our results to the even-to-odd transition, i.e., for the case where
there is (on average) one 
 electron less on the dot.
In Sec.~\ref{secSpinInverter}, we explain a mechanism for a spin inverter
device which inverts the spin-polarization of the current passing through
 two dots coupled in series in the presence of a homogeneous magnetic field.
In Sec.~\ref{secPumping}, we discuss how spin-dependent tunneling
 can be used to pump a current through a system  in the absence
 of a bias, where the ESR field provides the required energy.
In Sec.~\ref{secRotESR}, we consider rotating ESR fields which allows us
to obtain the exact time evolution of the dot-states
 and their decay rates.
In Sec.~\ref{secCotunneling}, the cotunneling current through
 the quantum dot away from the sequential tunneling peak is discussed.
We show that the $T_2$ time can also be accessed in this regime.
Invoking spin-polarized leads,  a read-out procedure for the dot-spin
 is proposed and analyzed in Sec.~\ref{secReadOut}, where counting statistics, noise,
 and the Fano factor are  calculated, which allow us then to estimate
 the measurement time. 
In Sec.~\ref{secRabiOsc}, we discuss coherent Rabi oscillations of the
dot spin and their occurrence in the
 time-dependent current.
In Sec.~\ref{secPulsedESR}, we show that
Rabi oscillations can also be observed in
 the time-averaged current if pulsed ESR fields are applied.
In Sec.~\ref{secSTM}, we point out that our results
also apply to STM devices, and  we finally conclude in
Sec.~\ref{secDisc}.

                \section{Quantum Dot in ESR Field}
\label{secQdot}

\subsection{Model Hamiltonian}
\label{ssecModelH}

We  consider a quantum dot in the 
 Coulomb blockade regime,\cite{kouwenhoven}
 which has a spin-$\frac{1}{2}$ ground state.
The dot is assumed to be tunnel-coupled to two Fermi-liquid leads $l=1$, $2$,
 at chemical potentials $\mu_l$.
We start from the full Hamiltonian 
\begin{equation}
\label{eqnHFull}
  H = \Hlead + \Hdot + \Hrf(t) + H_T,
\end{equation}
 which describes leads, dot, ESR field,
 and the tunnel coupling between leads and dot, respectively.
For the leads we take 
 $\Hlead = \sum_{lk\sigma}\epsilon^{}_{lk}c^\dagger_{lk\sigma}c^{}_{lk\sigma}$,
 where $c^\dagger_{lk\sigma}$ creates an electron in lead $l$
 with orbital state $k$,
 spin $\sigma$, and energy $\epsilon_{lk}$.
We describe the coupling with the standard
 tunnel Hamiltonian 
\begin{equation}
  H_T = \sum_{lpk\sigma} t_{lp}^{\sigma}c^\dagger_{lk\sigma}d_{p\sigma} + {\rm h.c.}\,,
\end{equation}
 with tunneling amplitude $t_{lp}^\sigma$, and
 where $d^\dagger_{p\sigma}$ creates an electron on the dot in
 orbital state $p$.
In Eq.~(\ref{eqnHFull}), $\Hdot$ is time-independent and 
 includes charging and interaction energies of the electrons
 on the dot and coupling to a static magnetic field $B_z$
 in $z$ direction.
The dot-spin is coupled to a magnetic ESR field,
 $B_x(t)=B_x^0 \cos(\omegarf t)$, linearly oscillating in
 $x$ direction with frequency $\omegarf$, 
 thus $\Hrf=-\frac{1}{2}g\mubohr B_x(t)\,\sigma_x$.
Such an oscillating field produces Rabi spin-flips 
 when its frequency is tuned to resonance, $\omegarf=\Delta_z$,
 as shown below.
Then, the total Zeeman coupling of the dot-spin is
\begin{equation}
-\frac{1}{2} g\mubohr \,\vec{B}(t)\cdot\bsigma =
-\frac{1}{2} \Delta_z \sigma_z- \frac{1}{2} \HrfOffDiag \,\sigma_x,
\end{equation}
 with electron $g$ factor $g$,
 Bohr magneton $\mubohr$, and Pauli matrices $\sigma_i$.
We have defined  $\HrfConst = g\mubohr B_x^0$, and
 the Zeeman splitting $\Delta_z=g\mubohr B_z$.
Ideally, we assume that the Zeeman splitting of the leads
 $\Delta_z^{\rm leads}$ is different from 
 $\Delta_z$, and
 $\Delta_z^{\rm leads}\ll \efermi$, where $\efermi$ is the Fermi energy,
 such that the effects of the fields $B_z$ and $B_x(t)$
 on the leads are negligible (see below). 
Such a situation can be achieved by using 
 materials of different $g$ factors \cite{Fiederling}
 and/or with local magnetic fields ($B_x$ or $B_z$).

We are neglecting photon assisted tunneling (PAT)
 processes,\cite{kouwenhoven,Brandes}
 in which oscillating electric potentials of the leads
 provide additional energy to electrons tunneling onto the dot.
We note that PAT contributions to the current 
 can be distinguished from ESR effects
 since the former contributions do not show resonant behavior as a function
 of $B_z$ and/or $\omegarf$, and they lead to several satellite peaks
 instead of one as for ESR effects (see below).
Further,
 if one avoids electrical rf components 
 parallel to the current, i.e., along the axis lead-dot-lead,
 no potential oscillations are produced,
 and thus PAT effects are excluded.
Finally,
 electric rf fields can be avoided altogether,
 using a setup as in Ref.~\onlinecite{Mooij}.
There, 
 the oscillating current induced in a superconducting wire
 (via an rf source) generates only a magnetic rf component in
 the near-field region,\cite{HarmansPrivate}
 with an the electric component that is negligibly small
 for $\omegarf\ll\omega_{\rm p}$,
 where $\omega_{\rm p}$ is the plasma frequency.

\begin{figure}
\centerline{\psfig{file=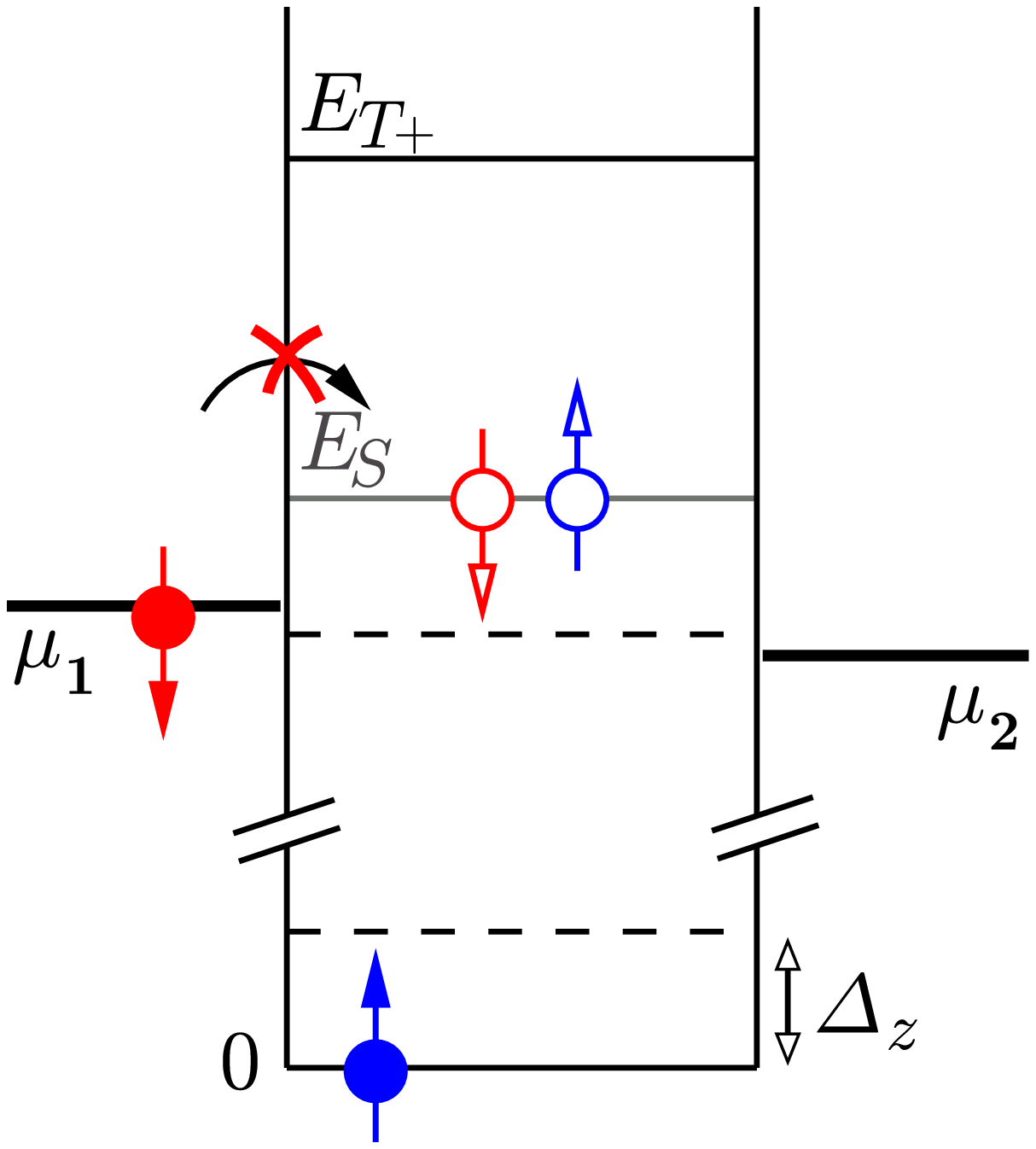,height=3.5cm}\hspace{3mm}\psfig{file=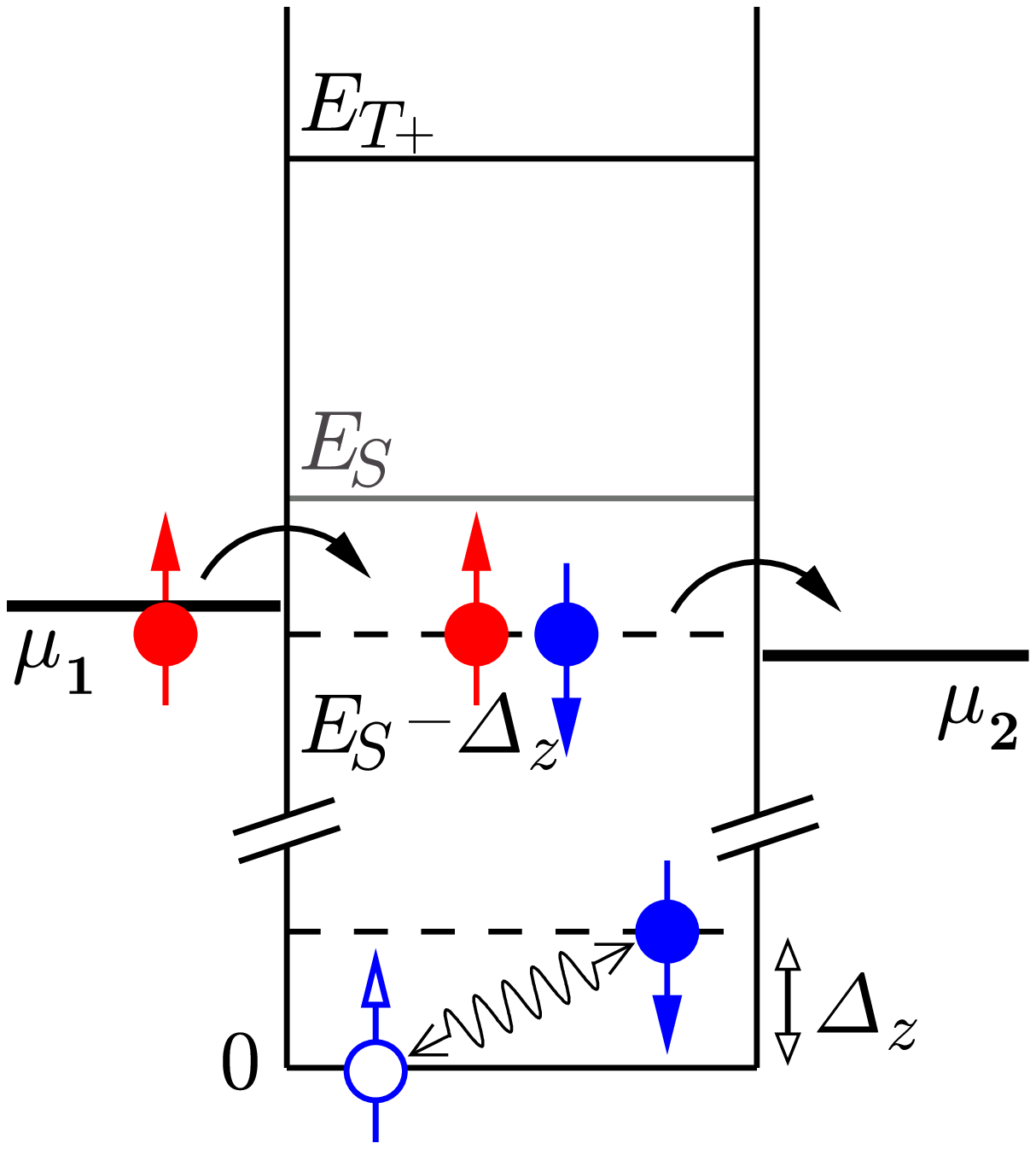,height=3.5cm}}
\caption[Dot]{
\label{figAcDot}
Quantum dot coupled to (unpolarized) leads $l=1$, $2$
 with chemical potentials $\mu_l$.
The sequential tunneling regime 
 $\ES > \mu_1 > \ES-\Delta_z > \mu_2$ (for $\Eu\!=\!0$)
 shown here corresponds to the satellite peak in the sequential
 tunneling current, 
 cf. Sec.~\ref{ssecSpinSatellite} and \ref{ssecMeasureTtwo}
 and Figs.~\ref{figCurrentVG} and \ref{figCurrentTT}.
Here, $\ES$ ($\ETp$) are the singlet (triplet) levels
 and the Zeeman splitting is $\Delta_z=g\mubohr B_z>kT$.
(a) If the dot is initially in the spin ground state $\spup$,
 sequential tunneling is blocked by energy conservation.
(b) If the dot-spin is excited by an ESR field (Rabi flip),
 spin up electrons can tunnel from lead 1 onto the dot,
 forming a singlet.
Then, spin up or down electrons can tunnel into lead 2.
}
\end{figure}

\subsection{Dot Spectrum and Energetics}
\label{ssecDotSpectrum}

The electronic states of the quantum dot can be assumed as follows.
For an odd number $N$ of electrons on a dot with antiferromagnetic filling,
 the dot has a spin-$\frac{1}{2}$ ground state.
The topmost (excess) electron can
 be either in the spin ground state, $\spup$, ($\sigma_z$ eigenstate)
 or in the excited state, $\spdown$ (see Fig. \ref{figAcDot}).
This assumption is automatically satisfied if $N=1$.
Otherwise, to obtain antiferromagnetic filling,
 Hund's rule must not apply.
This can be achieved by breaking the orbital degeneracy on the dot,
 e.g., by using asymmetrically shaped dots 
 or an appropriate magnetic field $B_z$.\cite{taruchaKouwenhoven}
For an additional electron on the dot,
 we assume for $N+1$ the ground state
 to be the singlet $\ket{S} = (\spupdown-\spdownup)/\sqrt{2}$,
 i.e., the triplet state $\ket{T_+}=\spupup$ has higher energy,
 which again can be achieved by tuning $B_z$.\cite{taruchaKouwenhoven}
The energy $E_m$ of the dot, including charging energy,
 is defined by $\Hdot \ket{m} = E_m \ket{m}$.

We shall give a brief overview of the energetics involved
 in tunneling through quantum dots 
 in the Coulomb blockade regime\cite{kouwenhoven}
 and in the presence of the Zeeman splitting and an ESR field.
For simplicity, we assume that there is no
 electron-electron interaction on the dot apart from the
 classical charging effect.
(Our work is not restricted to such an assumption,
 since we only require a spin-$\frac{1}{2}$ ground state
 and a large enough singlet-triplet spacing on the dot.)
The total ground state energy of a dot with antiferromagnetic
 filling is 
\begin{equation}
\label{eqnGroundStateEnergy}
U(N)=\sum_{k=1}^N \varepsilon_k^{\sigma} + E_C^N \,,
\end{equation}
 for $N$ electrons on the dot.
Here, the single particle energy of the $k$th electron,
 $\varepsilon_k^\sigma = \varepsilon_k + (-1)^k\Delta_z/2$,
 contains orbital and Zeeman energy contributions.
The charging energy is $E_C^N=(Ne-Q_G)^2/2C$, 
 with gate charge $Q_G$, and dot capacitance $C$.
It is convenient to define the chemical potential of the dot,
 $\kmudot(N+1) = U(N+1)-U(N)$,
 which is the energy required
 for an electron of lead $l$ to tunnel
 onto the dot, which contains $N$ electrons initially,
 i.e., tunneling onto the dot occurs for $\mu_l>\kmudot$.\cite{AdditionEnergy}
In the Coulomb blockade regime, $kT\ll e^2/C$ ($k$: Boltzmann constant), 
 no sequential tunneling current flows through the dot
 if the chemical potentials of dot and leads are such that
 $\kmudot(N)<\mu_1,\,\mu_2<\kmudot(N+1)$.
However, in the sequential tunneling regime $\mu_1>\kmudot(N+1)>\mu_2$,
 single electrons tunnel from lead 1 onto the dot
 and then on into lead 2,
 producing a sequential tunneling current.

In the presence of an ESR field, these concepts must
 be extended.
Excitations of the dot states must be taken into account, 
 since now the energy of the dot changes in time due to $B_x(t)$.
A full analytical description of the current flow is derived in the following
 sections based on a time dependent master equation.
Here, we just intend to give a qualitative picture to
 provide some intuition for the underlying physical mechanism
 (it will not be needed later on).
We define a time-dependent chemical potential of the dot, 
 given as the energy required to add an electron at time $t$.
We consider the two chemical potentials 
 $\kmudot^\sigma$ for initial spin-$\frac{1}{2}$ dot-state $\ket{\sigma}$, i.e.,
 $\ESu=\kmudot^\uparrow(N+1)=E_S-E_{\uparrow}$, and
 $\ESd=\kmudot^\uparrow(N+1)=E_S-E_{\downarrow}$,
 which simplify to $\ESu=E_S$, and $\ESd=E_S-\Delta_z$, respectively, 
 for $E_\uparrow=0$.
Note that the $\kmudot^\sigma$ is \textit{lowered}
 if the dot is excited into state $\spdown$,
 since the Zeeman energy $\Delta_z$ has already been provided 
 by a Rabi spin flip due to the ESR field.
Therefore, we can identify the regime $\ESu>\mu_1>\ESd>\mu_2$, 
 where a sequential tunneling current
 will flow through the dot only after exciting the dot-spin by a spin flip
 (see Fig.~\ref{figAcDot}).
In other words, the dot can be opened and closed via the ESR field,
 which thus allows to modulate the current.
This (dynamical) dependence of the current on the dot-spin
 can be exploited to measure the $T_2$ time and the Rabi oscillations
 of the dot-spin,\cite{EL}
 as we will explain in detail in the following.

\subsection{Systematic Treatment of Sequential Tunneling}
\label{ssecDotLeadCoupling}

The electronic states on a quantum dot
 interact with their environment (heat bath),
 in particular with the Fermi leads, which provide and take up electrons. 
The state of the combined system, dot and environment, 
 is given by the full density matrix $\rho(t)$.
The states of interest are the electronic states on the dot,
 described by the reduced density matrix of the dot, $\rhoSys = \TrB \rho$.
Here, $\TrB\!$ is the trace taken over the leads (environment),
 averaging over the (unobserved) degrees of freedom of the environment.
The diagonal elements $\rho_n=\bra{n}\rhoSys\ket{n}$ of
 the density matrix of the dot
 describe the occupation probabilities of the dot levels, 
 with $\Hdot\ket{n}=E_n\ket{n}$.
The off-diagonal elements $\rho_{nm}=\bra{n}\rhoSys\ket{m}=\rho_{mn}^*$
 describe the coherence and the phase of superpositions of dot-states.

The tunnel coupling $H_T$ between leads and dot is switched on at $t=0$.
Prior to this, the dot and leads are assumed to be uncorrelated
 such that the full initial density matrix factorizes as $\rho(0)=\rSNot\rBNot$,
 where $\rBNot$ is the density matrix of the leads in thermal equilibrium
 at $\mu_{1,\,2}$, and at temperature $T$.
Next we derive the master equation for the reduced density matrix $\rhoSys$ 
 by making use of the superoperator formalism.\cite{FickSauermann}
In the following, we set $\hbar=1$.
Starting from the von Neumann equation $\dot{\rho}=-i\,[H,\,\rho]$
 for the full density matrix,
 and using standard manipulations,\cite{FickSauermann}
 one finds the time evolution of the reduced density matrix
\begin{eqnarray}
\label{eqnRhoDotMemory}
\rhoDotSys(t) &=&  -i\, [\Hdot + \Hrf(t),\, \rhoSys(t)] 
\nonumber \\ && \quad
             -\int_{0}^t dt' M(t,\, t')\rhoSys(t'),
\\
\label{eqnKernelMFull}
M(t,\, t') &=& \TrB \, \LV 
  \left({\cal T}e^{-i \int_{t'}^t dt'' \, Q L(t'') } \right) 
 \LV \rBNot,
\end{eqnarray}
with time-ordering ${\cal T}$ and
 the Liouville operators defined by $L(t) X= [H(t),\,X]$, $\LV X= [H_T,\,X]$,
 and equivalently for $\Ldot$, $\Llead$, and $\Lrf(t)$.
The projectors are defined as $Q=1\!-\!P$, and $P X = \rBNot \TrB X$.
The kernel $M$ [Eq.~(\ref{eqnKernelMFull})] is a superoperator
 describing processes involving
 tunneling of electrons to and from the leads.
We consider here only sequential tunneling processes,
 and refer for a discussion of cotunneling contributions to
 Secs.~\ref{ssecCTcontributions} and \ref{secCotunneling}.
Thus, we work in Born approximation by
 retaining only the terms in lowest order of $\LV$,
 i.e., we replace $L$ by $L_0=L-\LV$ in Eq.~(\ref{eqnKernelMFull}).
For further evaluation of $M$, 
 it is self-consistent (see below) to neglect
 the effect of the ESR field, $\Lrf(t)$, i.e.,
 we replace $L_0$ by $\Ldot+\Llead$ in $M$.
This removes explicitly the time dependence of $M$,
 making it time translation invariant, $M(t,\, t')=M(t-t')$.
We find that $M(\tau)$ decays on a time scale $\tau_c \sim 1/kT$,
 i.e., the correlations induced in the leads by $H_T$ decay rapidly.
Since this decay is typically much faster than the Rabi flips produced by the ESR field,
 $\tau_c\ll 1/\HrfConst$,
 we may indeed neglect the contribution of $\Lrf(t)$ to $M$.
With these approximations, Eq.~(\ref{eqnRhoDotMemory}) becomes
 in the interaction picture
\begin{eqnarray}
\label{eqnRhoDotMemoryBorn}
\rhoDotSysIA(t) &=& -i \Lrf^{\rm I}(t) \rhoSysIA(t) 
             -\int_{0}^t \!\!d\tau\: M^{\rm I}(\tau)\rhoSysIA(t-\tau).
\end{eqnarray}
The rapid decay of $M(\tau)$ also
 justifies the Markovian assumption
 that the system has no memory about its past,
 i.e., that $\rhoDotSys(t)$ depends only on $\rhoSys(t)$
 and not on $\rhoSys(t-\tau)$.
This approximation is performed in the interaction picture,
 to keep track of the
 dynamical phase of the off-diagonal elements of $\rhoSys$.
Systematically we proceed as follows.
Since the integrand in Eq.~(\ref{eqnRhoDotMemoryBorn})
 only contributes for small $\tau$,
 we may expand the integrand in $\tau$,
 $M(\tau)\rhoSysIA(t-\tau)=M(\tau)[\rhoSysIA(t)-\tau\rhoDotSysIA(t)+O(\tau^2)]$.
We then replace $\rhoDotSysIA(t)$ in the integrand by using 
 Eq.~(\ref{eqnRhoDotMemoryBorn}) iteratively.
However, since $M(\tau)\sim O(\LV^2)$,
 we can neglect the part of $\rhoDotSysIA(t)$ which is $O(\LV^2)$,
 since it corresponds to a higher order term in our Born approximation.
The remaining part of $\rhoDotSysIA(t)$ results from $\Lrf$, 
 which can also be disregarded since, in the integrand,
 the ESR field only acts on the time scale $\tau_c\ll1/\HrfConst$.
We then extend the upper integration limit
 in Eq.~(\ref{eqnRhoDotMemoryBorn}) to $\infty$,
 with negligible contributions
 due to the decay of $M(\tau)$.
Therefore, the second term in  Eq.~(\ref{eqnRhoDotMemoryBorn}) becomes
 $-\big\{\int_{0}^\infty \!\!d\tau \, M^{\rm I}(\tau) \big\}\rhoSysIA(t)$.
Next, we evaluate the matrix elements
 $M_{bc|nm}=\bra{b}\Big(M\ket{n}\bra{m}\Big)\ket{c}$
 explicitly in the interaction picture,
 which yields\cite{Secular}
\begin{eqnarray}
\label{eqnMelements}
&&{}\!\!\!\!\!\!\!\!\!
-\int_0^\infty\!d\tau \: M_{bc|nm}^{\rm I}(\tau) = 
 \delta_{b c} \, \delta_{n m}  \bigg(W_{cn} - \delta_{bn} \sum_k W_{kn}\bigg)
\nonumber \\  
&&
 -(1-\delta_{nm}) \, \delta_{b n} \, \delta_{m c} 
 \Big[i\:\!\EStark_{nm}+
 \frac{1}{2} \sum_k  \left(W_{kn} + W_{km}\right) \Big],\quad\:\:
\end{eqnarray}
with the rates $W$ (see below) 
 and energy shifts $\EStark_{nm}$ (Stark shifts).
These shifts are small; e.g., the one
 between $\spdown$ and $\spup$ is given by
\begin{equation}
\label{eqnStarkShift}
\EStark_{\du}=
\frac{1}{2\pi} \sum_l \: {\cal P}\int_0^\infty \!\!d\epsilon\,\:
 f_l(\epsilon)\: \bigg(
 \frac{\gul{l}}{\epsilon-\ESd}-\frac{\gdl{l}}{\epsilon-\ESu}
 \bigg),
\end{equation}
 and similarly for 
 $\EStark_{S\downarrow}$ and $\EStark_{S\uparrow}$.
For $|\mu_l-\Delta_{S\sigma}|>kT$, the energy shift becomes

\begin{equation}
\EStark_{\du} = \sum_{l} \left(
  \frac{\gdl{l}}{2\pi} \log\left|\frac{\ESu}{\mu_l-\ESu}\right|
  -\frac{\gul{l}}{2\pi} \log\left|\frac{\ESd}{\mu_l-\ESd}\right|
 \right),
\end{equation}
 which, for $\gul{l}=\gdl{l}$, reduces to
  $\EStark_{\du} \approx \sum_l (\gamma_l/2\pi)
   \log \big[|\mu_l-\ESd|\big/|\mu_l-\ESu|\big]$,
 and thus to a small correction
 $|\EStark_{\du}| \lesssim \gamma \log(\Delta_z/kT)$,
 for $\Dm<\Delta_z$.

The sequential tunneling rates in Eq.~(\ref{eqnMelements}) are
\begin{eqnarray}
\label{eqnStRatesSd}
 \WSd = \sum_l \WSdl{l}, && \quad \WSdl{l} = \gul{l} \, f_l(\ESd)\,,
\\
\label{eqnStRatesdS}
 \WdS = \sum_l \WdSl{l}, && \quad \WdSl{l} = \gul{l} \, [1-f_l(\ESd)]\,,
\end{eqnarray}
 with the Fermi function $f_l(\ESd)=\big[1+e^{(\ESd-\mu_l)/kT}\big]^{-1}$
 of lead $l$.
The rates $\WSu$, $\WuS$, $\WSul{l}$, and $\WuSl{l}$ are defined analogously
 as functions of $\gdl{l}$ and $f_l(\ESu)$.
The transition rates,

\begin{eqnarray}
\label{eqnStRateZeroT}
 & \displaystyle
  \gul{l} = 2\pi \nu_{\uparrow}   \big|t_{l}^\uparrow\big|^2\,,\qquad
  \gdl{l} = 2\pi \nu_{\downarrow} \big|t_{l}^\downarrow\big|^2 \,,
\end{eqnarray}
 consist of (possibly) spin-dependent
 density of states $\nu_{\uparrow,\downarrow}$ at the Fermi energy
 and tunneling amplitude $t_{l}^{\uparrow,\downarrow}$.
(Spin-dependent density of states are considered
 in Sec.~\ref{secReadOut} for spin read-out.)
For later convenience, we define for $\sigma=\uparrow,\, \downarrow$

\begin{equation}
\gammaTZeroAvg^{\sigma}  = (\gammaTZero^{\sigma}_1 + \gammaTZero^{\sigma}_2)/2\,\,,
 \qquad
 \gAvg = (\guAvg+\gdAvg)/2\,.
\end{equation}

\subsection{Master Equation}
\label{ssecMasterEquation}

So far we have considered only coupling to 
 an environment consisting of Fermi leads.
However, the electronic dot states are affected also
 by intrinsic degrees of freedom 
 such as hyperfine coupling, spin-orbit interaction, or spin-phonon coupling,
 which lead to intrinsic spin relaxation and decoherence.
Treating such couplings microscopically is beyond the present scope
 (see e.g., Ref.\ \onlinecite{KhaetskiiNazarov}).
Thus, we treat these couplings
 phenomenologically by introducing corresponding rates
 in the master equation.
First, the spin \textit{relaxation} rates $\Wud$ and $\Wdu$
 describe processes in which the dot-spin is flipped.
We can assume $\Wud\gg\Wdu$, for $\Delta_z>kT$
 (consistent with detailed balance, $\Wud/\Wdu = e^{\Delta_z/kT}$).
These relaxation processes 
 correspond
 to the phenomenological rate $1/T_1 = \Wud + \Wdu$,
 see also Sec.~\ref{ssecDecoherence}.
Second, the rate $1/T_2$
 describes the intrinsic \textit{decoherence} of the spin on the dot,
 which is present even in the absence of coupling to the leads.
This type of decoherence
 destroys the information about the relative phase in a superposition
 of $\spup$ and $\spdown$, 
 without changing the populations of the opposite spin states.
Formally, this leads to a decay of the off-diagonal matrix element $\rhodu$.
Including the decoherence contribution of $H_T$
 [Eqs.~(\ref{eqnMelements}), (\ref{eqnStRatesSd})],
 the total spin decoherence rate is
\begin{equation}
\label{eqnGammdu}
  \gammadu = \frac{\WSu+\WSd}{2} + \frac{1}{\:T_2},
\end{equation}
 i.e., electrons tunneling onto the dot further destroy spin coherence
 on the dot 
 (see Sec.~\ref{ssecDecoherence} for an interpretation).

With the above results, we obtain from Eq.~(\ref{eqnRhoDotMemory})
 the master equation of the dot,

\begin{eqnarray}
\rhoDotuu &=&
   - (\Wdu+\WSu)\, \rhouu
   + \Wud \,\rhodd
   + \WuS \,\rhoSS
\nonumber \\ & & - \HrfConst\,\HrfOsc \:{\rm Im}[\rhodu],
\label{eqnRhoDotuu}
\\
\rhoDotdd &=&
    \Wdu \,\rhouu
  - (\Wud+\WSd)\,\rhodd
  + \WdS \,\rhoSS
\nonumber \\ & & + \HrfConst\,\HrfOsc \:{\rm Im}[\rhodu],
\label{eqnRhoDotdd}
\\
\rhoDotSS &=&
   \WSu\,\rhouu +\WSd\,\rhodd
  -(\WuS+\WdS)\,\rhoSS   ,
\label{eqnRhoDotSS}
\\
\rhoDotdu &=&
  - i \Delta_z \rhodu
  + i \frac{\HrfConst}{2}\HrfOsc (\rhouu -\rhodd)
 - \gammadu \, \rhodu,\quad\:
\label{eqnRhoDotdu}
\\
\rhoDotSu &=&
  - i \ESu \rhoSu - \gammaSu \, \rhoSu  ,
\label{eqnRhoDotSu}
\\
\rhoDotSd &=&
  - i \ESd \rhoSd -  \gammaSd \, \rhoSd.
\label{eqnRhoDotSd}
\end{eqnarray}
Here, the time evolution of
 the matrix elements $\rho_{nm}=\bra{n}\rhoSys\ket{m}$ 
 of the density matrix of the dot
 is described for the states $\ket{n}=\spup$, $\spdown$, $\ket{S}$,
 e.g., for the diagonal element we write 
 $\rhouu = \bra{\uparrow\!}\rhoSys\spup$,
 and for the off-diagonal element, $\rhoSu = \bra{S}\rhoSys\spup$, etc.
The rate $\WP_{mn}$ describes transitions 
 from state $\ket{n}$ to $\ket{m}$.
Equations (\ref{eqnRhoDotuu})--(\ref{eqnRhoDotSS}) are 
 rate equations with gain and loss terms,
 up to the contributions from the ESR field.
Then, the population of, say, state $\spup$,
 is changed by $d\rhouu$ 
 after time $dt$ by the following contributions [Eq.~(\ref{eqnRhoDotuu})].
The population $\rhouu$ is increased
 when the dot is previously in state $\ket{S}$ (with probability $\rhoSS$),
 and a spin $\downarrow$ electron tunnels out of the dot
 with probability $\WuS \, dt$.
However, the population  $\rhouu$ is decreased 
 when the system was already in state $\spup$, and
 a spin $\downarrow$ electron tunnels onto the dot
 with probability $\WSu \, dt$.
The spin flip rates, $\Wud$ and $\Wdu$, enter Eq.~(\ref{eqnRhoDotuu})
 analogously.
In the absence of an ESR field,
 the off-diagonal elements [Eqs.~(\ref{eqnRhoDotdu})--(\ref{eqnRhoDotSd})]
 of the density matrix decouple from the diagonal ones
 and decay with the decoherence rates $\gammaP_{nm}=\gammaP_{mn}$.

In the presence of an ESR field, the diagonal
 [Eqs.~(\ref{eqnRhoDotuu}) and (\ref{eqnRhoDotdd})] and the off-diagonal
 [Eq.~(\ref{eqnRhoDotdu})] matrix elements become coupled by the term
 proportional to $\HrfConst$.
This coupling of populations ($\rhouu$ and $\rhodd$) and
 coherence ($\rhodu$)
 shows the coherent nature of Rabi spin-flips
 and
 makes it apparent that we are studying a resonant process,
 which requires that we take $\Hrf$ fully into account.

The current $I_2=\expect{dq/dt}$ from the dot into lead 2 
 is defined by the number of charges $dq$ that accumulate in lead 2 after
 time $dt$.
With probability $\rhoSS$, the dot is in state $\ket{S}$ and
 a charge will tunnel into lead 2 with probability $\big(\WuSl{2}+\WdSl{2}\big)\,dt$.
However, if the dot is in state $\spup$ or $\spdown$, a charge
 may tunnel from lead 2 onto the dot, reducing the number of charges
 in lead 2.
Thus, in total we obtain for the current in lead 2
\begin{equation}
\label{eqnSTCurrentFromRates}
  I_2 =  e\,(\WuSl{2}+\WdSl{2})\,\rhoSS - e\,\WSul{2}\rhouu -e\,\WSdl{2}\rhodd.
\end{equation}
The current in lead 1, $I_1$, is obtained analogously and is given
 by Eq.~(\ref{eqnSTCurrentFromRates}) 
 after changing sign and replacing the index 2 by 1.
We show in Sec.~\ref{secStationaryCurrent} that
 $I_1=I_2$ in the stationary limit, due to charge conservation.

Finally we note that Eqs.~(\ref{eqnRhoDotSu}) and (\ref{eqnRhoDotSd}),
 which describe a superposition of an odd and an even number
 of electrons on the dot,
 decouple from Eqs.~(\ref{eqnRhoDotuu})--(\ref{eqnRhoDotdu})
 and are thus not of relevance for our considerations.
Further, since the coupling to the leads is 
 switched on only at $t=0$,
 initially the number of particles on the dot is well
 defined.
Therefore $\rhoSu$ and $\rhoSd$ vanish at $t=0$ and
 at all later times,
 as seen from
 Eqs.~(\ref{eqnRhoDotSu}) and (\ref{eqnRhoDotSd}).
In particular, no superposition of a state with an even and 
 a state with an odd number
 of electrons on the dot is produced by the coupling to the leads,
 since this would require a coherent superposition 
 of corresponding states in the leads,
 however, for times larger than $\tau_c$ 
 (which is typically the case), we can safely
 neglect any coherence in the Fermi liquid leads.

\subsection{Decoherence and Measurement Process}
\label{ssecDecoherence}

We elucidate the connection between spin decoherence and measurement,
first
in the absence of leads and ESR field.
We consider a
coherent superposition $\alpha\spup+\beta\spdown$
as the initial state of the dot.
This pure state corresponds to the reduced density matrix
$\rhouu(0) = |\alpha|^2$,  $\rhodd(0) = |\beta|^2$, and
$\rhodu(0) = \alpha\beta^*$,
and the master equation
contains only the rates $\Wud$, $\Wud$, and $\gammadu=1/T_2$.
The off-diagonal terms $\rhodu=\rho_{\ud}^*$,
decay with the decoherence time $T_2$,
$\rhodu(t) = e^{-t/T_2 -i t \Delta_z}\, \rhodu(0)$,
while the diagonal terms (occupation probabilities)
decay with the  spin relaxation time
$T_1=(\Wud+\Wdu)^{-1}$, $\rhodd(t) = \rhoddEq+e^{-t/T_1}\,
[\rhodd(0)-\rhoddEq]$,
toward their stationary value
$\rhoddEq=\Wdu/(\Wud+\Wdu)$, and $\rhouu=1-\rhodd$.
In total, for $T_2<T_1$, we can picture the decay of
 $\rhoSys$ as
\begin{equation}
\label{eqnRhoDecay}
\bigg( \begin{array}{cc}
   |\alpha|^2 & \alpha\beta^* \\ \alpha^*\beta & |\beta|^2
\end{array}\bigg)
\stackrel{\displaystyle T_2}{\longrightarrow}
\bigg(\begin{array}{cc}
   |\alpha|^2 &0\\ 0& |\beta|^2 \end{array}\bigg)
\stackrel{\displaystyle T_1}{\longrightarrow}
\bigg(\begin{array}{cc}
   \rhouuEq &0\\ 0& \rhoddEq \end{array}\bigg)\:,
\end{equation}
i.e.,  the off-diagonal terms vanish first
on the timescale $T_2$, and
then the diagonal ones equilibrate on the timescale $T_1$.

As shown in Sec.~\ref{ssecDotLeadCoupling},
when electrons  tunnel onto the dot, 
the decoherence rate $\gammadu$  [Eq.~(\ref{eqnGammdu})] and thus the decay of the off-diagonal
elements
is increased
further.
We note now the formal equivalence to the quantum measurement
process (in the $\sigma_z$ basis),
where the dot-spin is projected onto $\spup$ or $\spdown$,
and thus the off-diagonal matrix elements vanish.
This projection can be understood as a decoherence process.
Conversely, we can consider the decoherence due to tunneling
as a measurement performed by the tunneling electrons.
We note that this process is a \textit{weak} measurement in the following
sense.
The electrons in the leads attempt to tunnel on the dot,
but only with small probability $\propto W_{S\sigma}$ are
these attempts successful.
Thus, the current $I$, which carries away the
information of the dot state to the observer, is formed by these
successful electrons,
while the unsuccessful electrons are not detected.
Another way to say this is that a given electron from the lead
has only a small probability $\propto W_{S\sigma}$ to
``measure'' (i.e., decohere) the dot state.

\subsection{Cotunneling Contribution to the Sequential Tunneling Regime}
\label{ssecCTcontributions}

We work in the sequential tunneling regime, 
 defined by $\mu_1\!>\!\ESd\!>\!\mu_2$.
One can see that higher order---cotunneling---contributions can be
 neglected\cite{kouwenhoven,Recher}
 for $\gamma_l<\Delta_z,\,kT$, the regime of interest here.
Most importantly, the cotunneling contributions to $\gammadu$
 are of the order $\gamma_l^2/\Delta_z$ (see Sec.~\ref{secCotunneling}), 
 i.e., they are suppressed compared to the sequential tunneling
 contributions by a factor of $\gamma_l/\Delta_z$
 ($\approx 5\times10^{-5}$ for the parameters of Fig.~\ref{figCurrentTT}).
Formally, the cotunneling contributions to the master equation
 can be absorbed into $T_1$ and $T_2$.
For a discussion of cotunneling currents away from the sequential tunneling
 resonance see Sec.~\ref{secCotunneling}.

                     \section{Stationary current}
\label{secStationaryCurrent}
We now consider the stationary current $I$ in the presence of a
 continuous wave (cw) ESR field.
Therefore we calculate the stationary solution $\rho(t\to\infty)$ of the 
 master equation [Eqs.\ (\ref{eqnRhoDotuu})--(\ref{eqnRhoDotSd})].
We will apply the rotating wave approximation (RWA),\cite{Blum}
 where only the leading frequency contributions
 of $\Hrf$ are retained.
Higher order contributions would include simultaneous absorption of
 two photons and emission of another photon.
In lowest order, only single photons can be absorbed or emitted,
 producing a spin flip on the dot.
 To perform this approximation, we write 
 $\HrfConst\HrfOsc = 
  \frac{1}{2}\HrfConst\,(e^{i\omegarf t}+e^{-i\omegarf t})$,
 i.e., we decompose the linearly oscillating magnetic field into
 a superposition of a clockwise and an anti-clockwise rotating field.
Integrating
 Eqs.~(\ref{eqnRhoDotuu}), (\ref{eqnRhoDotdd}), and (\ref{eqnRhoDotdu}), 
 one finds that for $\omegarf\approx\Delta_z$, 
 the anti-clockwise rotating field leads to rapidly oscillating terms
 in the integrands, which nearly average to zero.
Therefore, we retain only the clockwise rotating field, 
 which is given by the term proportional to $e^{i\omegarf t}$
 (see also Sec.~\ref{secRotESR}).
Note that since only one field component contributes,
 the field amplitude is halved.
This leads to the period $T_\Omega$ of one Rabi oscillation, 
\begin{equation}
 T_\Omega = \frac{4\pi}{\Delta_x}.
\end{equation}
The RWA is valid for 
 $\HrfConst$, $\gammadu$, $|\Delta_z-\omegarf|\ll\omegarf$,
 see e.g., Ref. \onlinecite{RWA},
 and is well justified for the parameters considered here.
In the stationary case and using the RWA,
 the dependence of $\rhouu$ and $\rhodd$ 
 [Eqs.~(\ref{eqnRhoDotuu}) and (\ref{eqnRhoDotdd})] 
 on $\rhodu$ is eliminated,
 leading to the effective spin-flip rate
\begin{eqnarray}
\label{eqnWomegarf}
\Womegarf   
= \frac{\HrfConst^2}{8}
   \frac{\gammadu}{(\omegarf - \Delta_z)^2 + \gammadu^2},
\end{eqnarray}
 which is a Lorentzian as function of $\omega$
 with maximum $\WoMax=\HrfConst^2/8\gammadu$
 at resonance $\omegarf=\Delta_z$.

Now it is straightforward to find the stationary solution of 
 the effective rate equations for
 $\rhouu$, $\rhodd$ and $\rhoSS$,
\begin{eqnarray}
\label{eqnRhoStationaryuu}
\rhouu &\,=\,& \eta\, \big[ \WuS\WSd + (\Wud+\Womegarf)\,(\WuS+\WdS) \big],
\quad\:{}
\\
\label{eqnRhoStationarydd}
\rhodd &\,=\,& \eta\, \big[ \WdS\WSu + (\Wdu+\Womegarf)\,(\WuS+\WdS) \big],
\\
\label{eqnRhoStationarySS}
\rhoSS &\,=\,& \eta\, \big[ \WSu\WSd + \WSu (\Wud+\Womegarf) 
\nonumber \\ && \quad + \WSd (\Wdu+\Womegarf)  \big],
\end{eqnarray}
where the normalization factor
 $\eta$ is such that
$\sum_n\rho_n=1$.
We see from Eqs.~(\ref{eqnRhoStationaryuu})-(\ref{eqnRhoStationarySS})
 that the effective spin flip rates are
 $\Wud+\Womegarf$, and $\Wdu+\Womegarf$,
 i.e., the ESR field flips up and down spin
 with equal rate $\Womegarf$.

We can now calculate the spin-$\uparrow$ polarized current in lead 2, 
 $I_2^\uparrow=e\,\WdSl{2}\,\rhoSS-e\,\WSdl{2}\,\rhodd$ 
 [cf.\ Eq.~(\ref{eqnSTCurrentFromRates})].
The result is displayed in Eq.~(\ref{eqnFullStatCurrent}) in
 the Appendix. 
The spin-$\downarrow$ polarized current, $I^{\downarrow}_2$, is obtained 
 from Eq.~(\ref{eqnFullStatCurrent}) by interchanging
 $\uparrow$ with  $\downarrow$ in the numerator
 (the denominator remains unaffected by such an interchange).
The currents in lead 1, $I_1^{\uparrow,\downarrow}$ are obtained
 from the formulas for $I_2^{\uparrow,\downarrow}$
 by changing sign and interchanging indices 1 with 2.
Note that generally $I_1^\uparrow \neq I_2^\uparrow$,
 since the ESR field generates spin flips on the dot, and thus
 the spin on the dot is not a conserved quantity.
However, the stationary charge current $I_l=\sum_\sigma I_l^\sigma$
 is the same in both leads, $I=I_1=I_2$,
 due to charge conservation.

\subsection{Spin Satellite Peak}
\label{ssecSpinSatellite}

In this subsection
 we discuss the stationary current $I$ through the dot,
 in particular, its behavior as function of $\mu=(\mu_1+\mu_2)/2$,
 or, equivalently, as function of the gate voltage $V_g$.
We will see that an additional sequential tunneling peak (satellite peak)
 will appear due to the ESR field.
Before explicit evaluation of the current,
 we briefly describe this
 situation in qualitative terms.
We assume a large Zeeman splitting, $\Delta_z>\Delta\mu,\, kT$,
 with applied bias $\Delta\mu=\mu_1-\mu_2>0$.
If the potentials are such that $\mu_1>\ESu>\mu_2$,
 i.e., the chemical
 potential of the dot (relative to the ground state $\spup$)
 is between the chemical potentials of the leads,
 the state on the dot changes between $\spup$ and $\ket{S}$
 due to sequential tunneling events, leading
 to the standard sequential tunneling peak in $I(\mu)$ 
 at $\mu\approx\ESu$.

\begin{figure} 
\centerline{\psfig{file=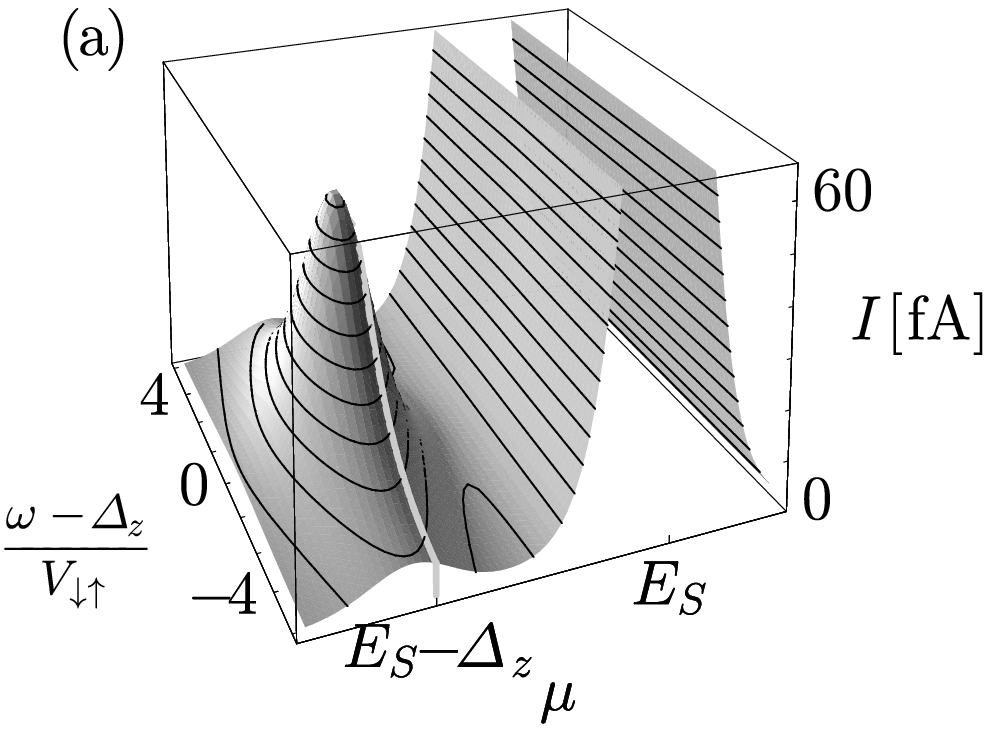,width=\figwidth}}
\vspace{4mm}
\centerline{\psfig{file=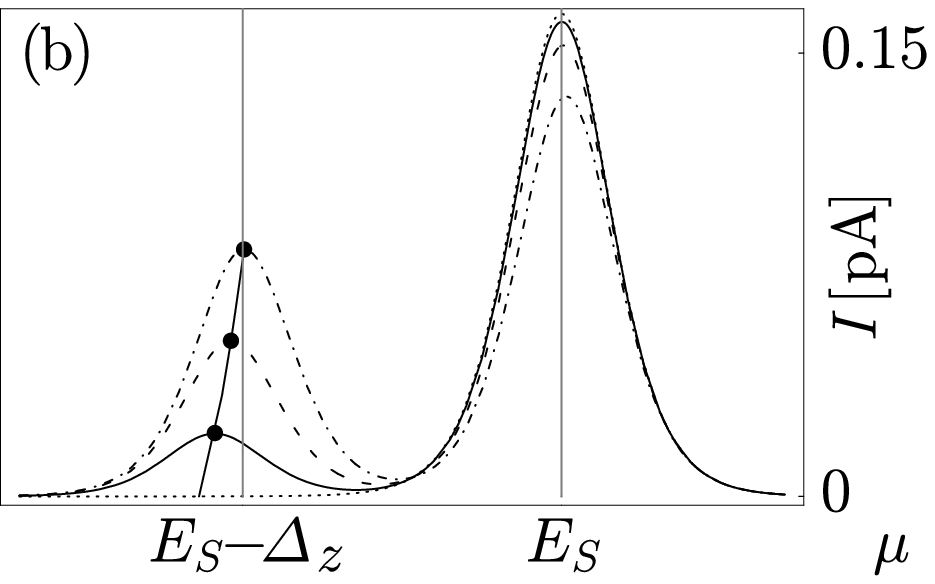,width=\figwidthSmall}}
\caption[Current]{
\label{figCurrentVG}
The stationary current $I$ [Eq.~(\ref{eqnFullStatCurrent})]
 vs.\ $\mu=(\mu_1+\mu_2)/2$
 and ESR frequency $\omegarf$.
We take $T=70\: {\rm mK}$, $\Delta\mu/e=6\:\mu{\rm V}$,
 $B_z = 0.5\: {\rm T}$,
 $g=2$,
 $T_1 = 1  \: \mu{\rm s}$,
 $T_2 = 100\: {\rm ns}$,
 $\gO=5\times10^6 \: {\rm s}^{-1}$, and
 $\gT=5\gO$,
 i.e., $\Delta_z=10kT$ and $\Dm=kT$.
The width of the sequential tunneling peaks in $I(\mu)$ is determined
 by the temperature, see Eq.~(\ref{eqnCurrentLinResp}).
(a) The current $I(\mu,\,\omegarf)$ shows
 a spin satellite peak near $\mu=\ES\!-\!\Delta_z$ (for $\Eu\!=\!0$)
 due to the ESR field.
Note that the spin satellite peak is slightly shifted from this position
 [see Eq.~(\ref{eqnPosSatellite})], 
 which is indicated by the line at $\ES\!-\!\Delta_z$
 (light gray line) in (a).
Here, $B_x^0=1.4\:{\rm G}$, i.e.,
 $\WoMax=\gO$ at resonance and $\mu=\ESd$.
(b)  The current $I(\mu)$ for
 $\Womegarf= 0$ (dotted), $\gO/5$ (solid), $\gO$ (dashed), $9\gO$ (dash-dotted).
The position of the spin satellite peak as function of $\Womegarf$
 is shown as black dots and the connecting solid line.
}
\end{figure}

However,
 we also have to consider the regime $\ESu>\mu_1>\ESd>\mu_2$,
 as shown in Fig.~\ref{figAcDot}.
Without ESR field,
 the dot relaxes into its ground state $\spup$ (since $\Wdu\ll\Wud$),
 and the sequential tunneling
 current through the dot is blocked
 since the chemical potential $\ESu$ of the dot
 is higher than those of the leads.
However, if an ESR field generates Rabi spin-flips (on the dot only), 
 the current flows through the dot involving the state $\spdown$,
 since $\ESd$
 is lower than $\mu_1$.
Therefore, a sequential tunneling current appears also
 for gate voltages $V_g$ corresponding to $\ESd$, 
 i.e., $I(\mu)$ 
 exhibits a spin satellite peak due to the ESR field at $\mu\approx\ESd$.
This new peak is shifted away from
 the main peak by $\Delta_z$ (Fig. \ref{figCurrentVG}).
The presence of such a satellite peak and its sensitivity to changes
 in $B_z$ allows identification of spin effects.\cite{PATFootnote}
Further, we note that
 via the position of the peak in $I(\omegarf)$, $I(B_z)$, or $I(\mu)$,
 the Zeeman splitting and also the $g$ factor of a single
 dot can be measured.
Such a measurement could provide a useful technique
 to study $g$ factor modulated materials, where
 the $g$ factor can be controlled by shifting
 the equilibrium position of the electrons in the dot
 from one layer to another by electrical gating.\cite{QCReview}
Note that measurement of the peak position would
 also allow to access the Stark shifts [Eq.~(\ref{eqnStarkShift})]. 

We consider now the analytic expression for the current $I$,
 as given in Eq.~(\ref{eqnFullStatCurrent}),
 for the regime of the spin satellite peak.
In this regime,
 $\ESu-\mu_1 =\ESd+\Delta_z-\mu_2-\Delta\mu 
  > \Delta_z-\Delta\mu \approx \Delta_z
  > kT$,
 and thus $f_l(\ESu)=0$, $\WSul{l}=0$, and $\WuSl{l}=\gdl{l}$.
For simplicity, we consider $\gls{l}{}=\gul{l}=\gdl{l}$ here
 (cf.\ Sec.~\ref{secPumping} for pumping due to $\gul{l}\neq\gdl{l}$).
The expression for the stationary current [Eq.~(\ref{eqnFullStatCurrent})]
 considerably simplifies to
\begin{eqnarray}
\label{eqnCurrentSatellite}
\nonumber
 I(\omegarf,\,\mu)&&=  2e \left( \Wdu+\Womegarf \right) 
 \gO\gT  [f_1(\ESd) - f_2(\ESd)] 
\\ \nonumber
 && \times
  \Big\{
  \left( 2\gAvg - \Wud - {\Womegarf} \right) 
  \left[\gO f_1(\ESd) + \gT f_2(\ESd) \right] \,
\\  && \quad
  + 4 \gAvg\left( \Wud+\Wdu+2{\Womegarf} \right) 
 \Big\}^{-1}
.
\end{eqnarray}
For a plot of $I$ vs $\omegarf$ and $\mu$ and some explanations of
 its characteristics, see Fig. \ref{figCurrentVG}.

\subsection{Spin Decoherence Time $T_2$}
\label{ssecMeasureTtwo}

\begin{figure} 
\centerline{\psfig{file=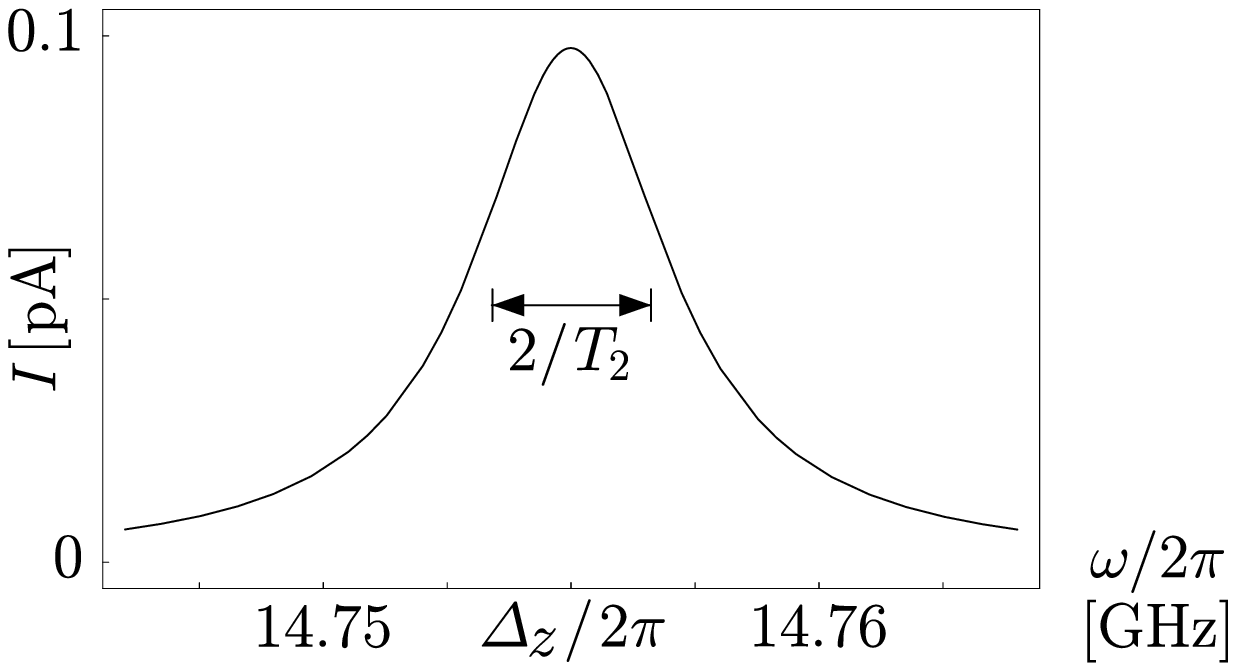,width=\figwidth}}
\caption[Current]{
\label{figCurrentTT}
The stationary current $I(\omegarf)$ [Eq.\ (\ref{eqnCurrentZeroT})]
 for $kT<\Dm$,
 $B_z = 0.5\: {\rm T}$,
 $B_x^0=0.45\: {\rm G}$,
 $T_1 = 1  \: \mu{\rm s}$,
 $T_2 = 100\: {\rm ns}$,
 $\gO=5\times10^6 \: {\rm s}^{-1}$, and
 $\gT=5\gO$, i.e., satisfying $\WoMax<\gO<1/T_2$.
Here, the linewidth gives a lower bound for
 the intrinsic spin decoherence time $T_2$
 (shown schematically by the arrow),
 while it becomes equal to $2/T_2$ for 
 $B_x^0=0.08\: {\rm G}$ and
 $ \WoMax\ll \gO=5\times10^5 \: {\rm s}^{-1}\ll 2/T_2$, where
 $I(\omegarf=\Delta_z) \approx 1.5\: {\rm fA}$.
}
\end{figure}

Around the spin satellite peak, it is possible to measure $\Womegarf$ via the current
 and thereby access the spin decoherence time  of the spin-$\frac{1}{2}$ state on the dot.
For this, we identify a regime where
 the Rabi spin-flips on the dot become the
 bottleneck for electron transport through the quantum dot
 such that the current becomes proportional
 to the spin-flip rate $\Womegarf$.
For $kT<\Delta\mu$ and
 $\WoMax < \mbox{max} \{\Wud,\, \gO \}$
 we obtain for the stationary current [Eq.~(\ref{eqnCurrentSatellite})],
\begin{eqnarray}
\label{eqnCurrentZeroT}
I(\omega) &=&
\frac{ 2 e \, \gO \gT \,(\Wdu+\Womegarf) }{
  \gO (\gO \!+\! \gT)
+     \Wud  (\gO \!+\! 2\gT ) },
\end{eqnarray}
 see Fig. \ref{figCurrentTT}.
We have used $\Wdu<\Wud$ here.
In the linear response regime, $kT>\Delta\mu$,
 and for $\WoMax < \mbox{max} \{\Wud,\, \gAvg f_1(\ESd+\Delta\mu/2) \}$,
 the current is
\begin{equation}
\label{eqnCurrentLinResp}
I(\omega) =
 \frac{e \, \gO \gT \,(\Wdu+\Womegarf) \: \Dm}
 {2(\gO+\gT) \, kT  \, h(T)} 
 \,\cosh^{-2}\!\left( \frac{\ESd-\mu}{2kT} \right).
\end{equation}
The current $I(\mu)$ shows the standard sequential tunneling peak
 shape, determined by the usual $\cosh$ dependence on temperature, 
 which is slightly modified by
 \begin{equation}
 h(T) = 2 \Wud + (2\gAvg-\!\Wud)\, f_1(\ESd\!+\!\Delta\mu/2).
 \end{equation}
Most importantly, the current $I(\omegarf)$ of the satellite peak
 [Eqs.\ (\ref{eqnCurrentZeroT}) and (\ref{eqnCurrentLinResp})]
 is proportional to the spin flip rate $\Womegarf$.
Thus, $I(\omegarf)$, 
 or equivalently $I(B_z)$,
 have a Lorentzian shape with resonance peak
  at $\omegarf=\Delta_z$ of width $2\gammadu$.
Since $\gammadu\geq 1/ T_2$, this width provides a lower bound on
 the \textit{intrinsic} spin decoherence time $T_2$
 of a single dot-spin.
For weak tunneling, $\gO\!<\!2/T_2$, this bound saturates,
 i.e., the width $2\gammadu$ becomes $2/T_2$.
Note that also Eq.\ (\ref{eqnCurrentDeltaMuZero}) (see below)
 shows resonant behavior,
 i.e., a lower bound for $T_2$ can also be measured via a current due to pumping.

We point out
 the similarity of our proposal to ESR spectroscopy,\cite{Abragam}
 where absorption or emission linewidths of the ESR field
 provide information on decoherence.
In contrast to these techniques,
 we are considering here linewidths in resonances of the current,
 which allows us to access even single spins,
 since very low currents can be measured accurately.

For Eqs. (\ref{eqnCurrentZeroT}) and (\ref{eqnCurrentLinResp}) we have
 assumed that $\Womegarf$ is small compared to the
 tunneling or the spin relaxation rates.
Therefore, we have neglected the contributions of $\Womegarf$
 in the denominator of these expressions.
 To take these contributions into account,
 we note that $\Womegarf/(\alpha+\Womegarf)$ as a function of $\omegarf$
 is still a Lorentzian,
 but with an increased width
 $w=2\gammadu\sqrt{1+\WoMax/\alpha}$.
Therefore, the current $I(\omegarf)$ has the linewidth
\begin{equation}
\label{eqnWidthZeroT}
 w = 2\gammadu \sqrt{1 +  \frac{\WoMax\, (3\gO \!+\! 4\gT )}{
    \gO (\gO \!+\! \gT)+ \Wud (\gO \!+\! 2\gT ) }} \:\:,
\end{equation}
for $kT<\Delta\mu$ [Eq. (\ref{eqnCurrentZeroT})], and
\begin{equation}
\label{eqnWidthLinResp}
 w = 2\gammadu \sqrt{1 + \WoMax\,[4-f_1(\ESd\!+\!\Delta\mu/2)] \big/ h(T) }\:\:,
\end{equation}
 for $kT>\Delta\mu$ [Eq. (\ref{eqnCurrentLinResp})].
Since the linewidth is increased by this correction, 
 the inverse linewidth is still a lower bound for $T_2$.

\subsection{Universal Conductance Ratio}
\label{ssecUniversalConductanceRatio}

\begin{figure} 
\centerline{\psfig{file=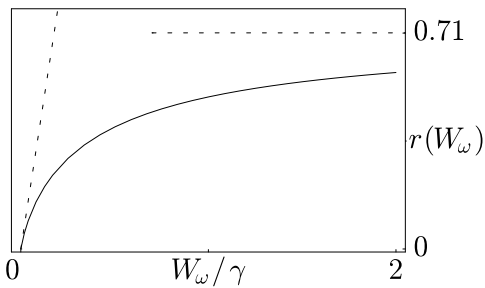,width=\figwidthSmall}}
\caption[Current]{
\label{figCurrentRatio}
The current ratio $r$ of the main and the satellite peak 
 as a function of the effective spin-flip rate $\Womegarf$
 [Eq.~(\ref{eqnConductanceRatio})].
The dashed line shows the saturation of $r$ for $\Womegarf\gg\gAvg$
 at the universal conductance ratio 
 $r_0 \approx 0.71$
 [Eq.~(\ref{eqnUniversalConductanceRatio})].
}
\end{figure}

For increasing $\Womegarf$, the satellite peak in the current $I(\mu)$
 increases while the main peak decreases,
 as shown in Fig.~\ref{figCurrentVG}(b).
Further, as function of $kT$, the peak is slightly shifted.
Explicitly, for $\gul{l}=\gdl{l}$, and $\Delta_z>\Delta\mu$, $kT$,
 we find from Eq.~(\ref{eqnCurrentSatellite})
 the position of the satellite peak
\begin{equation}
\label{eqnPosSatellite}
\mu_{\rm ESR} = \ESd -\frac{kT}{2} \log 
   \left\{ \frac{\Wud/2+\Wdu+3\Womegarf/2+\gAvg}{\Wud+\Wdu+2\Womegarf}\right\}
.
\end{equation}
The position of the main peak is 

\begin{equation}
\label{eqnPosMain}
\mu_0 = \ESu + \frac{kT}{2} \log 
   \left\{ \frac{\Wud+2\Wdu+3\Womegarf+2\gAvg}
            {\Wud+\Wdu+2\Womegarf+2\gAvg}\right\}
.
\end{equation}
An experimentally accessible quantity is the ratio
 of the two current peaks
 or, equivalently (for linear response $\Delta\mu<kT$),
 the ratio of the conductances
 $r(\Womegarf) =  I(\mu_{\rm ESR})\big/I(\mu_0) =  G(\mu_{\rm ESR})/G(\mu_0) $.
For this,
 we evaluate the stationary current at the gate voltages defined
 by Eqs.~(\ref{eqnPosSatellite}) and (\ref{eqnPosMain}),
 and find, for $\Delta\mu<kT$ and $\Wud<\Womegarf$,
\begin{equation}
\label{eqnConductanceRatio}
r(\Womegarf) = \frac{2\Womegarf
     \left(1+\sqrt{1+\frac{\Womegarf}{2\Womegarf+2\gAvg}}\right)^2 }
 {  4\sqrt{\Womegarf}\sqrt{3\Womegarf+2\gAvg}+
  (7\Womegarf+2\gAvg) }
,
\end{equation}
 see Fig. \ref{figCurrentRatio}.
On the one hand,
 for small spin-flip rates, $\Womegarf<\gAvg$,
 the ratio $r$ is $4\Womegarf/\gAvg$,
 i.e., at ESR resonance $r(B_x^0)=(g \mu_B B_x^0)^2/(2\gammadu\gAvg)$.
If the tunneling rates and field strengths are known,
 this provides a further method for measuring
 a lower bound of the single spin decoherence time.
On the other hand,
 this peak ratio [Eq. (\ref{eqnConductanceRatio})]
 can be used to measure the ratio $\Womegarf/\gAvg$,
 useful for estimating the additional peak broadening
 due to other limiting processes,
 as discussed in Sec.~\ref{ssecMeasureTtwo}, 
 cf.\ Eqs.\ (\ref{eqnWidthZeroT}) and (\ref{eqnWidthLinResp}).

It is noteworthy that this ratio saturates for $\Womegarf\gg\gAvg$ at
 the \textit{universal conductance ratio}
\begin{equation}
\label{eqnUniversalConductanceRatio}
r_0 = \frac{5+2\sqrt{6}}{7+4\sqrt{3}} \approx 0.71\:.
\end{equation}
For a larger bias, but still $\Delta\mu<\Delta_z$,
 and for $\Womegarf\gg\gAvg$, the ratio becomes
\begin{equation}
\label{eqnConductanceRatioDeltaMu}
r_0\left(\frac{\Dm}{kT}\right) = 
\frac{ 
    \left(\sqrt{3}+\sqrt{2}\:e^\frac{\Dm}{2kT}\right)^2\gO
   +\left(\sqrt{2}+\sqrt{3}\:e^\frac{\Dm}{2kT}\right)^2\gT
}{
    \left(2       +\sqrt{3}\:e^\frac{\Dm}{2kT}\right)^2\gO
   +\left(\sqrt{3}+2       \:e^\frac{\Dm}{2kT}\right)^2\gT
}.
\end{equation}
For $\gO=\gT$, the numerical value of $r_0$ remains $0.71$ for
 all values $\Delta\mu$.
Generally, $r_0$ is between $2/3$ (for $\gO\gg\gT$)
 and $3/4$ (for $\gO\ll\gT$),
 where $r_0$ takes these extremal values for $\Delta\mu>kT$.

Note that the current at the satellite peak is never
 larger than at the main peak.
This asymmetry is best explained in the limit $\Delta\mu>kT$, 
 when the ratio becomes
 $r_0(\infty) = (2\gO+3\gT)/(3\gO+4\gT)$.
Since $\Womegarf>\gAvg$,
 the Rabi spin flips equilibrate the populations $\rhouu$ and $\rhodd$.
Thus, the stationary populations of the states are
 $\rhoSS=\eta\,\Win$, and $\rhouu=\rhodd=\eta\,\Wout$,
 where $\eta=1/(\Win+2\Wout)$ 
 is a normalization factor,
 $\eta_{\rm ESR}$ at the satellite peak and $\eta_0$ at the main peak.
The rates $W_{\rm in(out)}$ include all
 processes of electrons tunneling into (out of) the dot.
Note that at the satellite peak, $\mu=\mu_{\rm ESR}$,
 a spin-up electron tunneling from lead 1 is the only process
 where an electron tunnels onto the dot, i.e., $\Win(\mu_{\rm ESR})=\gamma_1$,
 whereas at the main peak, $\mu=\mu_0$,
 the only tunnel process out of the dot is
 an electron with spin down into the right lead,
 i.e., $\Wout(\mu_0)=\gamma_2$.
At the satellite peak, both spin up and down electrons can tunnel 
 from the dot to lead 2, thus the current is given by
 $I(\mu_{\rm ESR}) = 2\gT\rhoSS=2\gO\gT\eta_{\rm ESR}$,
 with $\eta_{\rm ESR}=1/(3\gO+4\gT)$.
At the main peak, electrons can tunnel from lead 1 onto the dot, and the
 current is $I(\mu_0) = \gO(\rhouu+\rhodd)=2\gO\gT\eta_0$,
 with $\eta_0=1/(2\gO+3\gT)$.
Thus, the conductance ratio is given as $r_0=\eta_{\rm ESR}/\eta_0$,
 and we immediately obtain $r_0(\infty)$ in accordance with 
 Eq.~(\ref{eqnConductanceRatioDeltaMu}).
Therefore, the reason for $r_0<1$ is 
 that at the satellite peak three out of four tunnel processes
 contribute to $W_{\rm out}$, and thus $\eta_{\rm ESR}<\eta_0$,
 while only one contributes at the main peak.

\section{Even-to-odd sequential tunneling}
\label{secZeroOne}

Up to now we have considered sequential tunneling currents
 with odd-to-even transitions of the number of electrons on the dot.
Now we consider a different filling on the dot, 
 with even-to-odd transitions.
The state with $N$ even is $\ket{\SZ}$
 (involving different orbital states as for $\ket{S}$),
 and the states with $N+1$ are $\spup$ and $\spdown$.
This system can be described with the same formalism as before,
 but with the tunneling rates
 $\WZd=\sum_l \WZd^l\,{}$,  $\WdZ=\sum_l \WdZ^l\,{}$,
\begin{equation}
\label{eqnStRatesZeroElZd}
 \WZd^{l} =  \gdl{l} \, [1- \fZdl{l}]\,, 
 \qquad \WdZ^l =  \gdl{l} \, \fZdl{l}\,,
\end{equation}
and with $\WZu$, $\WuZ$, $\WZu^l$, and $\WuZ^l$ defined analogously.
The master equation of this system is given by Eqs.~(\ref{eqnRhoDotuu})-(\ref{eqnRhoDotSd})
 upon replacing the subscripts $S$ by $\SZ$.
Since $\WdZ$ describes an electron tunneling onto the dot, whereas
 $\WdS$ describes an electron tunneling out of the dot,
 the stationary current through the dot is given by
 Eq.~(\ref{eqnSTCurrentFromRates}) after changing its sign and replacing
 the subscripts, resulting in 
\begin{equation}
\label{eqnSTCurrentFromRatesZeroEl}
  I_2 = -e\,(\WuZ^2+\WdZ^2)\rho_{\SZ} +e\,\WZu^2\rhouu +e\,\WZd^2\rhodd\,.
\end{equation}
By comparing 
 Eqs.~(\ref{eqnStRatesSd}), (\ref{eqnStRatesdS})
 with (\ref{eqnStRatesZeroElZd}),
 and Eqs.~(\ref{eqnSTCurrentFromRates}) with (\ref{eqnSTCurrentFromRatesZeroEl}),
 we find that the formulas for the current are modified
 by the replacements
 $\fdl{l}\to[1-\fZdl{l}]$,
 $\gul{l}\to\gdl{l}$,
 $I_l^\uparrow\to -I_l^\downarrow$,
 and analogously for opposite spins.
For completeness, we give in 
 Appendix 
 the formula for 
 the stationary current $I^\downarrow_2$ [Eq.~(\ref{eqnFullStatCurrentZeroOne})],
 which is obtained by applying the above replacements 
 to Eq.~(\ref{eqnFullStatCurrent}).

\begin{figure}
\centerline{\psfig{file=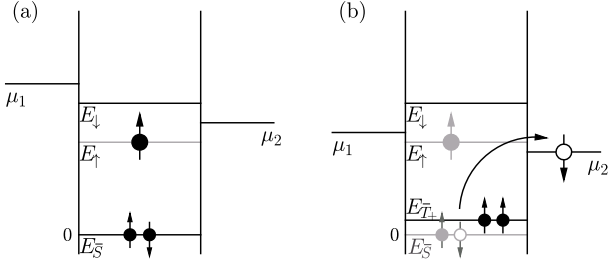,height=\figheight}
 \vspace{2mm}}
\caption[Dot]{
\label{figDotZeroOne}
(a) Setup for measuring $T_2$, with $\mu_1>\Ed>\mu_2$
 (for $E_{\SZ}=0$).
A lower lying state occupied by a singlet
 (corresponding to state $\ket{\SZ}$),
 illustrates the antiferromagnetic filling of the dot.
(b) Dot which should act as spin filter,
 allowing only spin $\uparrow$ to pass.
However, in the setup (b), the
 singlet-triplet spacing $E_{\bar{T}_+}-\EZ$ is too small
 compared to $\Dm=\mu_1-\mu_2$.
Here, if the initial dot state is $\spup$ (shown in gray),
 an electron with spin $\downarrow$
 from a lower-lying state can tunnel onto the right
 dot, leaving a triplet on the dot (black),
 thus the spin filter does not operate properly.
This problem disappears if the number of electrons on the dot
 can be reduced down to zero.
}
\end{figure}

In Sec. \ref{ssecMeasureTtwo} we have identified
 the regime of the spin satellite peak, which can be used to measure
 the decoherence time $T_2$.
For the setup considered here,
 an analogous regime is $\mu_1 > \EdZ > \mu_2 > \EuZ$,
 see Fig.~\ref{figDotZeroOne}(a).
The current at the spin satellite peak is then given
 by Eqs.~(\ref{eqnCurrentZeroT}) and (\ref{eqnCurrentLinResp}) in the corresponding regimes,
 after interchanging $\gO$ with $\gT$,
 replacing $f_1\to(1-f_1)$, and $\ESd\to\EdZ$.

For antiferromagnetic filling of the dot,
 one can use particle-hole symmetry to show that
 the two cases, odd-to-even and even-to-odd transitions,
 are equivalent.
Indeed, the tunneling from, say,
 a spin $\uparrow$ electron from the dot into the lead, $\spup\to\ket{\SZ}$,
 can be regarded as a spin $\uparrow$ hole which
 tunnels from the lead onto the dot, 
 which was initially occupied by a spin $\downarrow$ hole and now
 forms a hole singlet, i.e., $\ket{\!\downarrow_h}\to\ket{S_h}$.
With this picture in mind, above modifications become obvious.

\section{Spin Inverter}
\label{secSpinInverter}

\begin{figure}
\centerline{\psfig{file=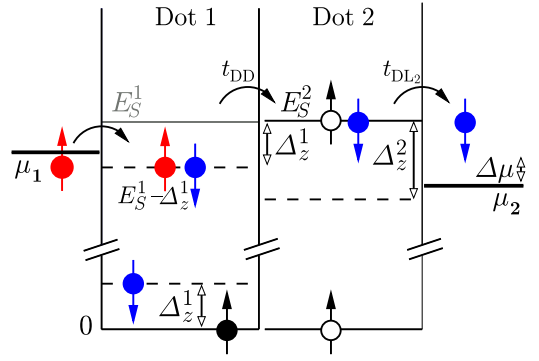,height=5cm}} 
\caption[Pump]{
\label{figSpinInverter}
Spin inverter setup, where
 the ESR field generates spin flips on dot 1,
 and the (additional) dot 2 acts as a spin filter, 
 allowing only spin $\downarrow$ electrons to tunnel into lead 2.
We consider the regime
 $|\tDD|<|\tDL|$,
 $\ES^1\approx\ES^2$, $\Delta_z^1\not\approx\Delta_z^2$,
 and $\ES^i > \mu_i > \ES^i-\Delta_z^i$, for $i=1,2$.
The allowed transition sequence is schematically given by
 \textQdot
 \edot{\uparrow}\qddot{\uparrow}{\uparrow}  
 $\stackrel{\mbox{\tiny ESR}}{\longrightarrow}$
 \edot{\uparrow}\qddot{\downarrow}{\uparrow} $\to$
                \qddot{\ud}{\uparrow}        $\to$
                \qddot{\uparrow}{\ud}      $\leftrightarrow$
                \qddot{\uparrow}{\uparrow}\edot{\downarrow} (see text),
 where ``$\leftrightarrow$'' means a coherent tunneling process.
}
\end{figure}

In this section we describe a setup with which spin-dependent tunneling, 
 $\gdl{l}\neq\gul{l}$, can be achieved.
Alternatively,  
 spin-polarized leads (see Sec.~\ref{secReadOut} for details)
 or spin-dependent tunneling barriers could be used.
This setup, shown in Fig.~\ref{figSpinInverter}, consist of two dots, 
 ``dot 1'' and ``dot 2'',
 which are coupled in series with inter-dot tunneling amplitude $\tDD$.
Dot 2 acts as a spin filter\cite{Recher}
 and is coupled to the lead 2 with tunneling amplitude $\tDL$.
We write the Zeeman splitting $\Delta_z^d$,
 the energy $E_n^d$ of state $\ket{n}$,
 and the chemical potential $\ESs{\sigma}^d$ with an index for dot $d=1,\,2$.
We assume that dot 2 remains unaffected by the ESR field,
 which can be achieved e.g., by applying $B_x$ and/or $B_z$ locally
 or with different $g$ factors for dot 1 and dot 2.
This assumption is taken into account by choosing
 $\Delta_z^1\not\approx\Delta_z^2$.

\subsection{Spin Filter}
 \label{ssecSpinFilter} 
We briefly review the concept of using a quantum dot
 as spin filter,\cite{Recher}
 as it is important for the description
 of the spin inverter.
If the dot is initially in state $\spup$, only a
 spin $\downarrow$ electron can tunnel onto the dot,
 forming a singlet.
Most importantly,
 the Zeeman splitting in the dot 
 should be such that $\Delta_z > \ESd-\mu_2$.
This ensures proper operation of the spin filter:
 because of energy conservation only the electron with spin $\downarrow$
 can tunnel from the dot to the lead,
 leaving the dot always in state $\spup$ after an electron has passed.
Therefore, the sequential tunneling current is spin $\downarrow$ polarized.
There is a small spin-$\uparrow$ cotunneling current,
 however, which is suppressed by a factor\cite{Recher} 
 $\gAvg\max\{kT,\,\Dm\}/(\ETp-\ES)^2$.
Note that for efficient spin filtering,
 it is favorable to have the singlet state $\ket{S}$ 
 as ground state with an even number of electrons on the dot,
 since the denominator of the suppression factor can become large,
 i.e., $\ETp-\ES>\Delta_z$.
Otherwise, if the triplet state $\ket{T_+}=\spupup$ is the 
 ground state, 
 only spin-$\uparrow$ sequential tunneling current can flow through
 the dot.
However, the spin-$\downarrow$ cotunneling current 
 involves the triplet state $\ket{T_0}=(\spupdown+\spdownup)/\sqrt{2}$,
 and the suppression factor is given by $\gAvg\max\{kT,\,\Dm\}/(\Delta_z)^2$,
 i.e., the cotunneling current is not suppressed efficiently.\cite{ABFootnote}

\subsection{Implementation of Spin Inverter}
 \label{ssecSpinInverterFunction}

For implementations of the spin inverter,
 the Zeeman splitting in dot 2
 should be such that $\Delta_z^2 > \ESd^1-\mu_2$,
 ensuring that dot 2 acts as a spin filter.
The coupling of dot 2 to the lead shall be strong
 such that electrons escape rapidly from dot 2 into lead 2.
This leads to resonant tunneling with resonance width 
 $\Gamma_2=2\pi \nu_\downarrow |\tDL|^2$.
We require $\Gamma_2 < \ESu^2-\mu_2$,
 i.e., that the broadened level of dot 2 is above $\mu_2$.
This excludes contributions from
 electrons tunneling from lead 2 onto dot 2,
 as shown in Ref.\ \onlinecite{AndreevQD}.

We calculate the rates $\hgu$ and $\hgd$
 for tunneling from dot 1 via dot 2 into lead 2
 in a $T$-matrix approach.\cite{Merzbacher,AndreevQD}
We use the tunnel Hamiltonian $H_T = \HDD + \HDL$,
 where $\HDD$ describes tunneling from dot 1 to dot 2
 and $\HDL$ from dot 2 to lead 2.
The transition rates are
$W_{fi} = 2\pi \biglb| \bra{f} T(\varepsilon_i)
  \ket{i} \bigrb|^2 \delta(\varepsilon_f-\varepsilon_i)$,
 where lead 2 is initially at equilibrium
 and with the $T$ matrix
\begin{equation}
T(\varepsilon_i)= \lim_{\eta\to+0}\, H_T \sum_{n=0}^\infty
  \left( \frac{1}{\varepsilon_i + i\eta - \Hdot-\Hlead} H_T \right)^n.
\label{eqnTmatrix}
\end{equation}
We take the leading order in $\HDD$ and sum
 up the contributions from all orders in $\HDL$.
We then integrate over the final states in lead 2
 and obtain the  Breit-Wigner transition rate
 of an electron with spin $\downarrow$
 to tunnel from dot 1 to lead 2 via the resonant level $\ES^2$ of dot 2,
\begin{equation}
\label{eqnResonantTunneling}
\hgd = 
   \frac{ |\tDD|^2 \Gamma_2}
     { (\ESu^1 - \ESu^2\big)^2 + \big(\Gamma_2/2\big)^2 }.
\end{equation}
In the spin filter regime considered here,
 dot 2 is always in state $\spup$.
Thus, tunneling of an electron with spin $\uparrow$ 
 would involve the triplet level $\ETp$ on dot 2,
 which is out of resonance,
 and thus $\hgu$ is suppressed to zero
 (up to cotunneling contributions, see Sec.~\ref{ssecCTcontributions}).
The state of dot 1 and the current through the setup is again described by
 the master equation [Eqs.\ (\ref{eqnRhoDotuu})--(\ref{eqnRhoDotSd})]
 with the tunneling rates $\WSdl{2}=\WdSl{2}=\WSul{2}=0$ 
 and $\WuSl{2}=\hgd$.
Thus, we can use all previous results for one dot in Sec.~\ref{ssecSpinSatellite},
 but with $\gdT\to\hat{\gamma}^\downarrow$, $\guT\to0$,
 and $f_2(\ESu)=0$.
Note that even for zero bias $\Delta\mu=0$, a pumping current flows 
 from lead 1 via the dots 1 and 2 to lead 2,
 see Eq.\ (\ref{eqnCurrentDeltaMuZero}) and Sec.~\ref{secPumping}.
We point out that this setup, see Fig. \ref{figSpinInverter},
 acts as a \textit{spin inverter}, i.e.,
 only spin $\uparrow$ electrons are taken as input (lead 1),
 while the output (lead 2) consists of 
 spin $\downarrow$ electrons.
In particular, the spin inverter
 does not require a change in the direction of the external magnetic field.

\section{Pumping}
\label{secPumping}

The ESR field provides energy to the system by
 exciting the spin state on the dot.
When the dot is initially in the excited state $\spdown$, 
 a spin up electron can tunnel onto the dot,
 followed by the spin down tunneling out of the dot.
In total, the Zeeman energy $\Delta_z$ is gained.
This energy input can be exploited to 
 induce a current through the dot, 
 even at zero bias $\Delta\mu=0$.
However, to obtain a directed current,
 the spin symmetry between lead 1 and 2 must be broken.
This can be achieved by 
 spin-dependent tunneling, $\gdl{l}\neq\gul{l}$,
 e.g., produced with a double-dot, see Sec.~\ref{secSpinInverter}.
At the spin satellite peak and for zero bias, i.e., $f_1=f_2$,
 there is a finite current [Eq.~(\ref{eqnCurrentSatelliteSpinDep})]
 due to ``pumping'' \cite{PP} by the ESR source,
\begin{eqnarray}
\label{eqnCurrentDeltaMuZero}
&&I(\omegarf)= e (\Wdu\!+\!\Womegarf)(\guO\gdT - \gdO\guT) f_1(\ESd)
\Big[
 2\guAvg f_1(\ESd) 
\nonumber \\ && \quad\times
  (2\gdAvg  \!-\! \Wud\!-\!\Womegarf)
 + 4\gAvg(\Wud\!+\!\Wdu\!+\!2\Womegarf)\Big]^{-1}.
\end{eqnarray}
Here, ${\rm sgn}(\guO\gdT - \gdO\guT)$
 determines the direction of the current.
Note that for spin-independent tunneling, $\gdl{l}=\gul{l}$,
 and the pumping current vanishes.

                       \section{Rotating ESR fields}
\label{secRotESR}

It is interesting to study \textit{rotating} magnetic fields
 in addition to linearly oscillating fields as studied above.
With rotating fields, it is possible to 
 calculate the time evolution of the density matrix
 of the dot exactly.
In particular, the stationary solution of the master equation 
 is obtained in a controlled approach
 and no rotating wave approximation is necessary.
However, rotating fields are experimentally more difficult to produce
 than linearly oscillating fields.

We consider a clockwise rotating field, described by 
\begin{equation}
\Hrf = -\frac{1}{4}\,\HrfRotConst\,
   [\sigma_x\cos(\omegarf t)-\sigma_y\sin(\omegarf t)],
\end{equation}
 where $\HrfRotConst = 2 g \mubohr B_\perp^0$.
Thus, for $\HrfConst=\HrfRotConst$ 
 we have chosen the amplitude of the
 rotating field to be only half the amplitude of the 
 linearly oscillating field,
 since both lead to the same effective spin flip rate $\Womegarf$.
Using Eq.~(\ref{eqnRhoDotMemory}) we immediately obtain the master equation,
 which is given by Eqs.~(\ref{eqnRhoDotuu})--(\ref{eqnRhoDotSd}) after the 
 following replacements.
The last terms in Eqs.~(\ref{eqnRhoDotuu}) and (\ref{eqnRhoDotdd}) become
 $\mp (\HrfRotConst/2)\, {\rm Im}\big[ e^{i\omegarf t} \rhodu \big]$, 
 respectively.
Equation (\ref{eqnRhoDotdu}) is replaced by

\begin{equation}
\rhoDotdu =
  - i \Delta_z \rhodu
  + i \frac{\HrfRotConst}{4} e^{-i\omegarf t} (\rhouu -\rhodd)
 - \gammadu \, \rhodu  \,.
\label{eqnRotRhoDotdu}
\end{equation}
We transform to the rotating frame,
 $\spup_r = e^{i\omegarf t/2}\spup$, and
 $\spdown_r =e^{-i\omegarf t/2}\spdown$,
 such that $\rhodu = e^{-i\omegarf t}\rhodu^r$.
This transformation removes the time-dependence of the coefficients
 in the master equation,
 which we shall now write as $\brhoDotSys^r = {\cal M} \brhoSys^r$.
The equations for $\rhoDotSu$ and $\rhoDotSd$ decouple and
 we write the remaining part of the 
 superoperator ${\cal M}$ as matrix in the basis 
 $\{\rhouu^r,\, \rhodd^r,\, \rhoSS^r, {\rm Re}[\rhodu^r],  {\rm Im}[\rhodu^r]\}$,

\begin{widetext}

\begin{eqnarray}
\label{eqnMasterMatrix}
{\cal M} &=&
\left(
\begin{array}{*{4}{c@{\quad}}c}
-(\Wdu+\WSu)  & \Wud & \WuS &  0 & -\HrfRotConst/2
\\
\Wdu   & -(\Wud+\WSd) & \WdS & 0 &  \HrfRotConst/2
\\
\WSu &  \WSd & -(\WuS+\WdS)  &   0 &  0
\\
0 &  0  & 0 & -\gammadu & (\Delta_z-\omegarf)
\\ 
\HrfRotConst/4 &  -\HrfRotConst/4  & 0 & -(\Delta_z-\omegarf) & -\gammadu
\end{array}
\right).
\end{eqnarray}
\end{widetext}

\noindent
The master equation can now be solved exactly by calculating the 
 eigenvalues $\lambda_i$ of ${\cal M}$.
Since the total probability is conserved,
 $\sum_n \dot{\rho}_n = 0 = \sum_{nm} {\cal M}_{nm}\rho_m$,
 where $n$ is summed  over the diagonal elements, 
 and $m$ over diagonal and off-diagonal elements of $\brhoSys$.
By considering linearly independent initial conditions for $\brhoSys$,
 we see that $\sum_{n} {\cal M}_{nm}=0$, for every $m$.
Thus, adding up the rows in ${\cal M}$ for the diagonal elements of $\brhoSys$
 gives zero,
 which is satisfied explicitly by adding the first three rows 
 in Eq.~(\ref{eqnMasterMatrix}).
Therefore,
 ${\cal M}$ does not have full row rank
 and there is an eigenvalue $\lambda_0=0$
 with eigenspace describing the stationary solution.
The eigenvalues of ${\cal M}$ are

\begin{eqnarray}
\label{eqnEigenvalues}
&&\bigg\{0, -\gammadu, -3 W,
\\ \nonumber &&
 -\frac{1}{2}\bigg(\Sigma_W + \gammadu \pm 
   \sqrt{ (\Sigma_W-\gammadu)^2 - \HrfRotConst^2 } \bigg)
\bigg\}
,
\end{eqnarray}
with $\Sigma_W = W+\Wud+\Wdu$,
 and where we have considered $W=\WSu=\WSd=\WuS=\WdS$,
 and resonance $\Delta_z=\omegarf$
 for simplicity.
If all $\lambda_i$ are different,
 the time evolution of the density matrix is
 $\brhoSys(t) = \sum_i c_i e^{\lambda_i t} \brho_i$.\cite{FootnoteEigenvects}
The decay of the contribution of the eigenvectors $\brho_i$
 is exponential 
 and generally all decay rates $\lambda_i$ are involved.
Further, 
 we see from the last two eigenvalues in Eq.~(\ref{eqnEigenvalues})
 that the decay rates of $\brhoSys$ may be a nontrivial
 function of the rates involved in the master equation.
This should be kept in mind
 when one uses time dependent ensemble properties, i.e.,
 $\brhoSys(t)$, to
 measure intrinsic rates, e.g., $T_1$ and $T_2$.
We point out that the presence of very small decay rates
 does not necessarily prevent a decay of the initial conditions.
If, say, the tunneling rates are smaller than the spin relaxation rate,
 $W\ll \Wud$,
 it would be interesting to study a density matrix which is described
 as a linear combination of
 the eigenvector with eigenvalue $-3W$ [Eq.~(\ref{eqnEigenvalues})]
 and the stationary solution $\brho_0$, 
 i.e., $\brhoSys(t)=\brho_0 + c\, e^{-3Wt}\brho_{3W}$,
 which is independent of $\Wud$.
However, such an initial condition
 always contains contributions from state $\ket{S}$
 such that, in particular, it is not possible to construct an initial
 spin-$\frac{1}{2}$ state 
 which would decay only with the slow rate $3W$.

The (exact) stationary solution of the master equation can
 be readily obtained from Eq.~(\ref{eqnMasterMatrix}).
By eliminating $\rhodu^r$ from the coupled equations,
 we obtain the effective spin flip rate
\begin{eqnarray}
\label{eqnRotWomegarf}
\Womegarf   
= \frac{\HrfRotConst^2}{8}
   \frac{\gammadu}{(\omegarf - \Delta_z)^2 + \gammadu^2},
\end{eqnarray}
which is equivalent to Eq.~(\ref{eqnWomegarf}). 
Thus, all the results for the stationary currents
 from Sec.~\ref{secStationaryCurrent} apply 
 and are exact for the case of rotating magnetic fields.

                      \section{Cotunneling}
\label{secCotunneling}

\begin{figure} 
\centerline{\psfig{file=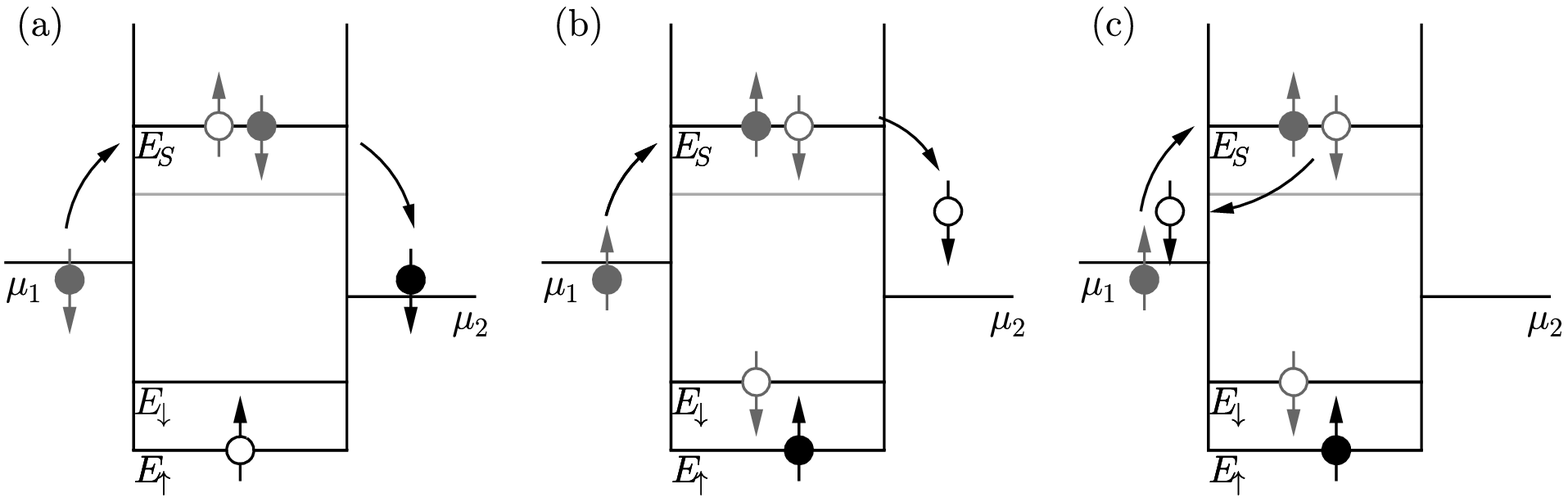,width=\figwidth}}
\caption[Cotunneling]{
\label{figCotunneling}
Cotunneling processes involving $\ket{S}$ for $\Delta_z>\Dm$.
(a) Elastic cotunneling. 
The cotunneling sequence
  \captionQdot   
  \edot{\downarrow}\qdot{\uparrow} $\to$
  \qdot{\uparrow\downarrow} $\to$
  \qdot{\uparrow}\edot{\downarrow},
 involving the virtual state $\ket{S}$ on the dot
 with virtual energy cost $\ESu-\mu_1$.
An equivalent process is possible
 when the initial and final dot state is $\spdown$,
 however, with a virtual energy cost reduced by $\Delta_z$.
These elastic cotunneling processes contribute to transport and to
 spin decoherence,
 while they do not contribute to spin relaxation (i.e., $T_1$).
(b) Inelastic cotunneling from lead 1 into lead 2 via the sequence
  \captionQdot   
  \edot{\uparrow}\qdot{\downarrow} $\to$
  \qdot{\uparrow\downarrow} $\to$
  \qdot{\uparrow}\edot{\downarrow}.
Note that
 tunneling of an electron from lead 2 into lead 1 is also possible,
 since the energy gain $\Delta_z$ from the dot relaxation is larger than
 the bias $\Dm$.
(c) Inelastic cotunneling, where only one lead is involved.
The process shown here leads to a particle-hole excitation in lead 1.
While it does not directly contribute to transport,
 it contributes to spin relaxation and spin decoherence of the dot.
}
\end{figure}

We now consider the cotunneling 
 regime\cite{averinnazarov,Schoeller,NoiseLong}
 $\ESu$, $\ESd>\mu_1$, $\mu_2\gg\Ed$, $\Eu$,
 where the number of electrons on the dot is odd,
 thus the state on the dot is described by $\spup$ and $\spdown$.
The leading order tunnel processes is
 now the tunneling of electrons
 from lead $l$ onto the dot, forming a virtual state $\ket{n}$, followed
 by tunneling into lead $l'$.
The spin state of the dot changes  $\sigma\to\sigma'$.
This process is called elastic cotunneling for $\sigma=\sigma'$
 and inelastic cotunneling for $\sigma\neq\sigma'$.
Note that in the absence of an ESR field,
 the dot relaxes into its spin ground state and
 no inelastic cotunneling processes, exciting the dot spin,
 occur for $\Dm<\Delta_z$.
However, if an ESR field is present, 
 the dot-spin can be excited by spin flips.
Then, inelastic cotunneling processes, which relax the dot-spin,
  can occur.
These processes either contribute to transport 
 or produce a particle-hole excitation in lead 1 or 2
 [see Fig.~\ref{figCotunneling}(b) and (c)].

These cotunneling rates are calculated in a ``golden rule'' approach,\cite{Recher}
 which is known to be consistent with a 
 microscopic derivation,\cite{NoiseLong}
\begin{equation}
\label{eqnCTRate}
W_{\sigma' \sigma}^{l'l}=
 2\pi \nu^2 \!\!\int \!\! d\epsilon \: 
  f_l(\epsilon)\,[1-f_l'(\epsilon\!-\! \Delta_{\sigma'\sigma})]\:
 \bigg| \sum_n \frac{t^{}_{l'\sigma'n} t^*_{l\sigma n}}{\Delta_{n\sigma}-\epsilon} \bigg|^2,
\end{equation}
where the possible spin-dependence of $\nu$ has been absorbed into $t$,
 and $\Delta_{\sigma'\sigma} = E_{\sigma'}-E_{\sigma}$ is
 the change of Zeeman energy on the dot, and $\Delta_{n\sigma}=E_n-E_\sigma$ is 
 the energy cost of the virtual intermediate state.
Here, $t_{l\sigma n}$ are the tunneling amplitudes,
 where $t_{l\downarrow S}=t^\uparrow_l$
 has already been introduced in Eq.~(\ref{eqnStRateZeroT}).
The cotunneling current through the dot can be calculated by
 summing up the contributing tunneling rates, as we have done
 for Eq.~(\ref{eqnSTCurrentFromRates}),
\begin{equation}
\label{eqnCTCurrentFromRates}
  I_{\rm CT}= e\sum_{\sigma\sigma'} 
  (W_{\sigma' \sigma}^{21}-W_{\sigma' \sigma}^{12})\, \rho_\sigma.
\end{equation}
We point out that by treating the cotunneling processes with golden
 rule rates, only classically allowed dot-states are considered.
Thus, the number of charges on the dot is fixed
 and no charge can temporarily accumulate as for sequential tunneling.
In particular, we have neglected quantum charge fluctuations
 on the dot.
Therefore, within our master equation approach for cotunneling,
 the charge currents in both leads are equal,
 $I_1(t)=I_2(t)$.
This equality is valid for ``coarse-grained'' expectation values of the current
 (and other physical observables).
In this approximation,
 one smoothens out the quantum fluctuations by averaging over the short-time behavior,
 i.e., one considers only the behavior on time scales larger than
 the lifetime $1/(\ESd-\mu)$ of the virtual states on the dot.
However, when the charge imbalance due to the virtual states 
 is taken into account in a microscopic treatment,
 one can find pronounced peaks in the noise $S(\omega)$
 for $|\omega|$ corresponding to the virtual energy cost,
 as it was shown in Ref. \onlinecite{DDot}.

The inelastic cotunneling provides spin relaxation processes in 
 addition to those contributing to $T_1$,
 totaling in $\WCTud = \Wud + \sum_{ll'} W_{\ud}^{l'l}$.
For processes with $l'=l$, particle-hole excitations are produced in lead $l$.
We are interested in the regime $\Delta\mu<\Delta_z$,
 where (inelastic) cotunneling does not excite the dot-spin, i.e., $\WCTdu = \Wdu$.
In analogy to Eq.~(\ref{eqnGammdu}), we take a phenomenological
 total spin decoherence rate
\begin{equation}
\gammaCTdu= \frac{1}{T_2} + \frac{1}{2}\sum_{ll'\sigma\sigma'} W_{\sigma' \sigma}^{l'l}\:,
\end{equation}
 where all spin relaxation and tunneling processes are taken into account.
The master equation for the dot in the cotunneling regime and in the 
 presence of a linearly polarized ESR field becomes
\begin{eqnarray}
\rhoDotuu &=&
   - \WCTdu \, \rhouu
   + \WCTud \,\rhodd
 - \HrfConst\,\HrfOsc \:{\rm Im}[\rhodu],
\label{eqnCTRhoDotuu}
\\
\rhoDotdd &=&
    \WCTdu \,\rhouu - \WCTud\,\rhodd
+ \HrfConst\,\HrfOsc \:{\rm Im}[\rhodu],
\label{eqnCTRhoDotdd}
\\
\rhoDotdu &=&
  - i \Delta_z \rhodu
  + i \frac{\HrfConst}{2}\HrfOsc (\rhouu -\rhodd)
 - \gammaCTdu  \rhodu   
. \qquad{}
\label{eqnCTRhoDotdu}
\end{eqnarray}
Note that away from the sequential tunneling regime,
 the master equation becomes much simpler while the 
 formulas for the rates are more involved.

For the time-averaged current we evaluate the stationary solution
 of the master equation 
 in the rotating wave approximation 
 (see Sec.~\ref{secStationaryCurrent}) for linearly
 or exactly (see Sec.~\ref{secRotESR}) for circularly polarized ESR fields.
This yields an effective spin-flip rate 
 $\Womegarf$ [Eqs.~(\ref{eqnWomegarf}) and (\ref{eqnRotWomegarf}), respectively] 
 and eliminates Eq.~(\ref{eqnCTRhoDotdu}).
We obtain 
\begin{equation}
\label{eqnCTrhoddStat} 
\rhodd = \frac{\Womegarf+\Wdu}{2\Womegarf + \Wdu + \Wud+ \sum_{ll'} W_{\ud}^{l'l}}
\end{equation}
and $\rhouu=1-\rhodd$.
We consider the case
 close to a sequential tunneling resonance (but still in the cotunneling regime),
 $\ESs{\sigma}-\mu_l<\ETp-\ES$,
 such that the virtual energy cost of an intermediate triplet
 state is much higher than that for a singlet state.
Since $(\ETp-E_{\sigma}-\mu)/(\ES-E_{\sigma}-\mu)<1$,
 with $\mu=(\mu_1+\mu_2)/2$,
 we have to consider only cotunneling processes involving state $\ket{S}$
 in Eq.~(\ref{eqnCTRate}).
For $\Delta\mu$, $kT < \ESd-\mu < \ETp-\ES$, 
 the relevant elastic rates are 
\begin{equation}
\label{eqnWCTEl}
 W_{\sigma\sigma}^{21} = \frac{\gO\gT}{2\pi} 
  \frac{\Dm}{(\Delta_{S\sigma}-\mu)^2}.
\end{equation}
The inelastic rates are,  for lead indices $l$, $l'=1$, 2,
\begin{eqnarray}
\label{eqnWCTInel}
 W_{\ud}^{l'l}&=&
   \frac{\gO\gT}{2\pi} 
  \frac{\Delta_z + (l'\!-\!l)\Dm}{(\ESd-\mu)(\ESd+\Delta_z-\mu)}
\\ &\approx&
\label{eqnWCTInelOffPeak}
 \frac{\Delta_z + (l'\!-\!l)\Dm}{\Dm} \: W_{\downdown}^{21},
\end{eqnarray}
where  Eq.~(\ref{eqnWCTInelOffPeak}) is valid for  $\Delta_z < \ESd-\mu$.
Note that for $\Dm<\Delta_z$ the inelastic rates can be much 
 larger (by a factor of $\Delta_z/\Dm$) than the elastic ones,
 while their contribution to the current,
 $W_{\ud}^{21}-W_{\ud}^{12}=2W_{\downdown}^{21}$,
 is of the same order as for the elastic rates.

For $\WoMax,\Wud<W_{\ud}^{21}$, 
 we obtain the cotunneling current from 
 Eqs.~(\ref{eqnCTCurrentFromRates}) and
 (\ref{eqnCTrhoddStat})--(\ref{eqnWCTInel}),
\begin{eqnarray}
\label{eqnCTcurrent}
I_{\rm CT} &=& 
\frac{e}{2\pi}\frac{\Dm\, \gO\gT}{(\ESu-\mu)^2}
\nonumber \\  && 
 + e\,\Womegarf \frac{\Dm}{4\,\Delta_z} 
     \bigg[3 - \frac{\ESd-\mu}{\ESu-\mu}+\frac{\Delta_z}{\ESd-\mu} 
             \bigg]
\\ &\approx&
\frac{e}{2\pi}\frac{\Dm\, \gO\gT}{(\ESu-\mu)^2}
 + e \Womegarf \frac{\Dm}{2\,\Delta_z}.
\end{eqnarray}
The first term in Eq.~(\ref{eqnCTcurrent}) results from 
 elastic cotunneling
 with spin ground state $\spup$ on the dot.
The second term represents the increased current 
 if the spin is flipped into state $\spdown$ before cotunneling occurs,
 since then both elastic and inelastic cotunneling processes contribute
 to the current.
The current $I_{\rm CT}$ is proportional to $\Womegarf$,
 up to a constant background,
 and thus shows, as a function of $\omega$,
 a resonant peak at $\omegarf=\Delta_z$ of width $2\gammadu$.
Thus, the intrinsic spin decoherence time $T_2$ is accessible
 in the cotunneling current as well as
 in the sequential tunneling (see Sec.~\ref{ssecMeasureTtwo}).
Generally, the cotunneling current is much smaller 
 than the sequential tunneling current,
 and thus it might seem more difficult to detect $T_2$ in 
 the cotunneling regime. 
However, since current and decoherence rate due to tunneling
 are proportional to $\gAvg^2$,
 the small currents can be compensated by choosing more transparent
 tunnel barriers, i.e., larger $\gAvg$.
Then, current and decoherence rate 
 in the cotunneling regime can become comparable to the
 sequential tunneling values given in Sec.~\ref{ssecMeasureTtwo}.
For illustration we give the following estimates.
For 
 $B_z = 1\:{\rm T}$, 
 $B^0_x = 2\:{\rm G}$,
 $g=2$,
 $\gO=\gT=5\times10^9\:{\rm s}^{-1}$,
 $T_1 = 1\:\mu{\rm s}$,
 $T_2 = 100\:{\rm ns}$,
 $\ESd-\mu=\Delta_z$, and
 $\Delta\mu = \Delta_z/5$,
 the cotunneling current as function of the ESR frequency $\omega$
 is $0.17\:{\rm pA}$ away from resonance and 
 exhibits a resonance peak of $I_{\rm CT}^{\rm max}=0.31\:{\rm pA}$,
 with half-width $\gammaCTdu=3.41\times10^7\:{\rm s}^{-1}$.

              \section{Spin Read-out with spin-polarized leads}
\label{secReadOut}

An electron spin on a quantum dot
 can be used as a single spin memory
 (or as a quantum bit for quantum computation\cite{QCReview}),
 if the spin state of the quantum dot can be measured.
It was shown that a quantum dot connected to
 fully spin-polarized leads, $\Delta_z^{\rm leads}>\efermi>\Delta_z$,
 can be used for reading the spin state of the quantum dot via the 
 charge current.\cite{Recher}
Such a situation  can be realized with magnetic
 semiconductors (with effective g-factors exceeding 100)\cite{Fiederling}
 or in the quantum Hall regime
 where spin-polarized edge states are coupled to a quantum
 dot.\cite{Sachrajda}
If the spin polarization in both leads is $\uparrow$, 
 no electron with spin $\downarrow$ 
 can be provided or taken by the leads (since $\nu_\downarrow=0$),
 and the rates $\WSu$ and $\WuS$ vanish.
Thus, if the dot is initially in state $\spup$, 
 no electron can tunnel onto the dot
 (the formation of the triplet is forbidden by energy conservation)
 and $I=0$, up to negligible cotunneling contributions.
However, if the dot is in state $\spdown$, a current can flow via
 the sequential tunneling transitions
  \textQdot
  \edot{\uparrow}\qdot{\downarrow} $\to$
  \qdot{\uparrow\downarrow} $\to$
  \qdot{\downarrow}\edot{\uparrow}.
Therefore, the initial spin state of the quantum dot can be 
 detected by measuring the current through the dot.
Note that for this read-out scheme, it is not necessary to have
 $\Delta_z>kT$ on the dot, the constraint of having 
 spin-polarized leads is already sufficiently strong.

In the stationary regime and for $\Delta_z>kT$,
 the current becomes blocked due to
 spin relaxation ($\Wud$).
However, this blocking can be removed
 by the ESR field producing spin flips on the dot
 (with rate $\Womegarf$).
For $\Womegarf<\Wud$, this competition leads
 again to a stationary current with resonant structure,
\begin{equation}
I(\omegarf)=e\,(\Wdu+\Womegarf)\, \frac{\gO\gT}{\gT\,\Wud + (\gO+\gT)\,\Wdu},
\end{equation} 
 from which $\gammadu$ (and 1/$T_2$) can be measured.
Note that the relaxation rate $\Wud$ is rather small, thus
 only small ESR fields can be used, which leads to small currents.

\subsection{Counting Statistics and Signal-to-Noise Ratio}
\label{ssecCounting}

We analyze now the time-dynamics of the read-out of a dot-spin
 via spin-polarized currents.
The goal is to obtain the full counting statistics and to
 characterize a measurement time $\tmeas$ for the spin read-out.
While we have considered only averaged currents so far,
 we now need to keep track of the number of
 electrons $q$ which have accumulated in lead 2
 since $t=0$.\cite{Jong}
The time evolution of $\rhoSys(q,\,t)$, now charge-dependent,
 is described by
 Eqs.\ (\ref{eqnRhoDotuu})--(\ref{eqnRhoDotSd}),
 but with replacements
 $\WdSl{2}\,\rhoSS(q)\to\WdSl{2}\,\rhoSS(q-1)$ in Eq.\ (\ref{eqnRhoDotdd}), and
 $\WSdl{2}\,\rhodd(q)\to\WSdl{2}\,\rhodd(q+1)$ in Eq.\ (\ref{eqnRhoDotSS}).
Next, we consider the distribution function
 $P_i(q,\,t) = \sum_n \rho_n(q,\,t)$
 that $q$ charges have accumulated 
 in lead 2 after time $t$ 
 when the dot was in state $\ket{i}$ at $t=0$.
For a meaningful measurement of the dot-spin,
 the spin flip times $\Wud^{-1}$, $\Wdu^{-1}$, and $1/\HrfConst$ must 
 be smaller than $\tmeas$ and are neglected.
Eqs.\ (\ref{eqnRhoDotuu})--(\ref{eqnRhoDotSd})
 then decouple except 
 Eqs.\ (\ref{eqnRhoDotdd}) and (\ref{eqnRhoDotSS}),
which we solve for $\rhouu=1$, and for $\rhodd=1$ at $t=0$.
 The general solution follows by linear combination.
First, 
 if the dot is initially in state $\spup$,
 no charges tunnel through the dot,
 and thus $P_\uparrow(q,\,t)=\delta_{q0}$.
Second,
 for the initial state $\spdown$,
 we consider $kT<\Delta\mu$
 and equal rates $\WSdl{1}=\WdSl{2}=W$.
We relabel the density matrix
 $\rhodd(q)\to\rho_{m=2q}$, and $\rhoSS(q)\to\rho_{m=2q+1}$,
 and Eqs.\ (\ref{eqnRhoDotdd}) and (\ref{eqnRhoDotSS}) become
\begin{equation} 
 \dot{\rho}_m = W (\rho_{m-1} - \rho_m),
\end{equation} 
 with solution
 $\rho_m(t) = (Wt)^m e^{-Wt} / m!$ (Poissonian distribution).
We obtain the counting statistics
\begin{equation}
P_\downarrow(q,\,t) = \frac{(Wt)^{2q} e^{-Wt}}{(2q)!} 
 \left(1+\frac{Wt}{2q+1}\right).
\end{equation}
Experimentally, $P_\downarrow(q,\,t)$ can be determined
 by time series measurements or by using an array of 
 independent dots (see Sec.~\ref{lblEnsembleAvg}).
The inverse signal-to-noise ratio is
 defined as the Fano factor,\cite{BB,Devoret}
 which we calculate as 
\begin{equation}
F_\downarrow(t)= \frac{\expect{\delta q(t)^2}}{\expect{q(t)}} = 
 \frac{1}{2} +   \frac{3 - 2e^{-2Wt}(4Wt+1) -  e^{-4Wt}}{
         4 \big(2Wt-1 + e^{-2Wt}\big)} ,
\end{equation}
 with $F_\downarrow$ decreasing monotonically
 from $F_\downarrow(0)=1$ to $F_\downarrow(t\to\infty)=\frac{1}{2}$.
Note that for dot-spin $\spup$,
 only weak cotunneling occurs
 with Fano factor $F_\uparrow=1$.\cite{NoiseLong}

If we are interested in the current and noise for long times $t>W^{-1}$, 
 we can follow the steps used in Ref.~\onlinecite{Schoen}.
We decouple the differential equations with respect to $q$ 
 by taking the inverse Fourier transform,
 $\rhoSys(k)=\sum_q e^{-ikq}\rhoSys(q)$.
Note that, for $k=0$, we recover the
 density matrix $\rhoSys=\rhoSys(k=0)$,
 where the accumulated charge is not taken into account.
The probability $P_\downarrow(q,\,t)$
 is approximated by a Gaussian wave packet in $q$-space with group velocity
 $I/e = \WSdl{1}\WdSl{2}/(\WSdl{1}+\WdSl{2})$, and width $\sqrt{2 F (I/e) t}$,
 and 
\begin{equation}
 F=\frac{(\WSdl{1})^2+(\WdSl{2})^2}{(\WSdl{1}+\WdSl{2})^2}
\end{equation}
 is the Fano factor.\cite{Schoen}
However, within this approximation, valid for $Wt>1$, we 
 cannot access the short time behavior where only a few electrons
 have tunneled through the dot,
 which is of importance for the read-out process considered here.

\subsection{Measurement Time}
\label{ssecMeasurementTime}

Using the counting statistics,
 we can now quantify the measurement efficiency.
If, after time $\tmeas$, some charges $q>0$ have tunneled through the dot,
 the initial state of the dot was $\spdown$
 with probability 1
 (assuming that single charges can be detected via an SET \cite{Devoret}).
However, if no charges were detected ($q=0$),
 the initial state of the spin memory was 
 $\spup$ with probability 
\begin{equation}
\label{eqnMeasureProb}
1-P_\downarrow(0,\,t) = 
  1-\frac{ \WSdl{1}e^{-\WdSl{2}t}-\WdSl{2}e^{-\WSdl{1}t} }{ \WSdl{1}-\WdSl{2} },
\end{equation} 
 which reduces to $1-e^{-Wt}(1+Wt)$,
 for equal rates. 
Thus, roughly speaking, we find that 
 $\tmeas \gtrsim 2W^{-1}$, as expected,
 while the Fano factor is $0.5<F_\downarrow\lesssim0.72$.
If, more generally, the threshold for detection is  at $m$ charges,
 $m\geq1$, Eq.~(\ref{eqnMeasureProb}) is replaced by
 $1-\sum_{q=0}^{m-1} P_\downarrow(q,\,t)$.

We insert now realistic numbers to obtain an estimate of the 
 fastest possible measurement time which can be achieved
 with this set-up.
For a fast spin read-out, the tunneling rates and the current
 trough the dot should be large,
 limited by the fact that the conductance
 of the dot should not exceed the single-channel conductance $e^2/h$.
In the linear response regime and for a small bias $\Dm/e$,
 the current is $I= e \guAvg \Dm/8kT < (\Dm/e)\times(e^2/h)$,
 for $\guO=\guT$.
Thus, the tunneling rates are limited by
 $\guAvg < 8 kT/h = 1.76\times 10^{11}\, (T/{\rm K})\: {\rm s}^{-1}$.
For $W=\guAvg=1.25\times10^{10}\:{\rm s}^{-1}$
 (corresponding to $kT<\Dm$ and a current $I=1\:{\rm nA}$),
 and $m=1$,
 the spin state can be determined with more than $95\%$ probability
 for a measurement time of $\tmeas=400\:{\rm ps}$, 
 and with more than $99.99\%$  probability for $\tmeas=1\:{\rm ns}$.\cite{ReadOutFootnote}

       \section{Rabi Oscillations of a single Spin in the time domain}
\label{secRabiOsc}

\subsection{Observing Rabi Oscillations via Current}
\label{ssecRabiOscCurrent}

\begin{figure} 
\centerline{\psfig{file=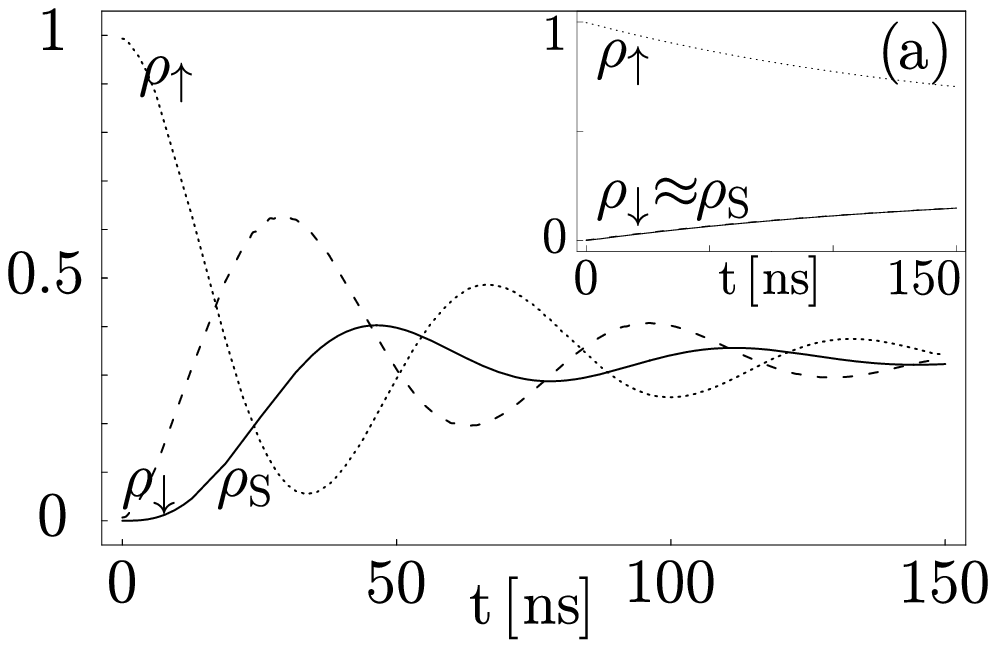,width=\figwidthSmall}}
\centerline{\psfig{file=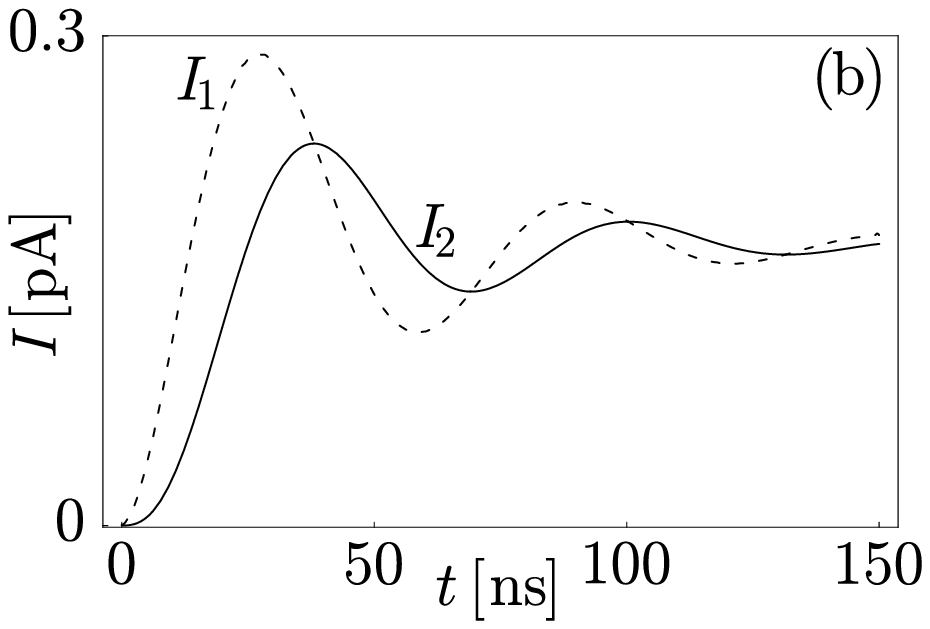,width=\figwidthSmall}}
\caption[Rabi oscillations]{
\label{figRabiOsc}
Rabi oscillations of the electron spin on the dot in the time domain.
We consider the regime at the spin satellite peak,
 $\ESu>\mu_1>\ESd>\mu_2$, (see Fig.~\ref{figAcDot})
 and take
 $T_1=1\: \mu{\rm s}$, 
 $T_2 = 300\: \rm{ns}$, 
 $\HrfConst = 5\WSd$ 
 (corresponding to $B_x^0=10\: \rm{G}$ for $g=2$),
 and $\rhouu=1$ at $t=0$.
During the time span shown here, less than 3 electrons 
 have tunneled through the dot on average.
Here, the spin decoherence 
 is dominated by the tunneling process, i.e., $\WSd\gg1/T_2$.
(a)
 Spin-polarized leads with the only non-vanishing tunnel rates
  $\WSd=\WdS=4\times10^7 \: \rm{s}^{-1}$.
The Rabi oscillations show up in
 $\rhouu$ (dotted), $\rhodd$ (dashed) and $\rhoSS$ (full line),
 which is directly visible in the current,
 since $I_1^\uparrow(t)\propto\rhodd$ and
 $I_2^\uparrow(t)\propto\rhoSS$, for $kT<\Dm$.
In the inset, we show the case of
 large tunneling, $\WSd=\WdS=10^9 \: \rm{s}^{-1}\gg \HrfConst$.
As a consequence of the Zeno effect (see Sec.~\ref{ssecZeno}),
 the Rabi oscillations are suppressed.
Further, $\rhodd$ and $\rhoSS$ are indistinguishable since
 $\spdown$ and $\ket{S}$ equilibrate rapidly due to the increased tunneling.
(b) The time-dependent currents in unpolarized leads,
   $I_1(t) = e\gO(\rhodd-\rhoSS)$ and $I_2(t) = 2 e\gT \rhoSS$,
 for 
   $kT<\Dm$, and
   $\gul{l}=\gdl{l}=4\times10^7 \: \rm{s}^{-1}$,
   for $l=1,2$.
}
\end{figure}

The ESR field generates coherent Rabi oscillations of
 the dot spin, leading to oscillations in $\rhoSys(t)$.
Since the time-dependent currents $I(t)$ in the leads are given by the
 populations $\rho_n(t)$
 [Eq.~(\ref{eqnSTCurrentFromRates})], 
 current measurements give access to these Rabi oscillations.
First, we consider a dot coupled to unpolarized leads
 in the regime of the spin satellite peak 
 (see Fig.~\ref{figAcDot} and Sec.~\ref{ssecSpinSatellite}).
For $kT<\Delta\mu$,
 the current in lead 2 is $I_2(t) = e (\guT+\gdT)\, \rhoSS(t)$,
 i.e., $\rhoSS$ is directly accessible via measurement 
 of $I_2(t)$.\cite{FootnoteChargeAccum}
Further, for $\guO=\gdO$, the current in lead 1 is
 $I_1(t)= e \gO (\rhodd-\rhoSS)$, 
 which gives access to $\rhodd(t)$, if the ratio $\gO/\gT$ is known.
We calculate the oscillations of $I_{1,\,2}(t)$ explicitly by numerical
 integration of the master equation
 [Eqs.\ (\ref{eqnRhoDotuu})--(\ref{eqnRhoDotdu})],
 see Fig.~\ref{figRabiOsc}(b).

The measurement of $\rhoSys$ can be refined
 by using the spin read-out setup
 with spin-polarized leads (Sec.~\ref{secReadOut}).
For $kT<\Delta\mu$,
 the current is $I_1(t)=I^\uparrow_1(t) = e \guO \rhodd(t)$ in lead 1, 
 and $I_2(t)=I^\uparrow_2(t) = e \guT \rhoSS(t)$ in lead 2.\cite{FootnoteChargeAccum}
Thus, the time-dependence of $\rhodd$ and $\rhoSS$
 (and also of $\rhouu=1-\rhodd-\rhoSS$)
 can be directly measured via the currents $I_{1,\,2}$,
 see Fig.~\ref{figRabiOsc}(a).

Note that the electrons which tunnel onto the dot decohere
 the spin state of the dot (see Sec.~\ref{ssecDecoherence}).
Thus, to observe Rabi oscillations in $I_{1,\,2}(t)$ experimentally,
 the Rabi frequency $\HrfConst$ 
 must be larger than the coupling to the leads $\WSd$,
 otherwise the strong decoherence
 (equivalent to a continuous measurement)
  suppresses the Rabi oscillations
 (Zeno effect, see Sec.~\ref{ssecZeno}).
Then, however, only very few electrons tunnel per Rabi oscillation 
 period through the dot.
 To overcome the limitations of such a weak current signal 
 and to obtain $I_{1,\,2}(t)$ experimentally,
 an ensemble average is required.

\label{lblEnsembleAvg}
There are two possibilities to obtain averages, namely
 using many dots or performing a time series measurement.
First,
 many independent dots can be measured simultaneously
 by arranging the dots in parallel to increase the total current.
For example, an array (ensemble) of dots and leads could be produced
 with standard techniques for defining nanostructures,
 or self-assembled or chemically synthesized dots could be placed
 within an insulating barrier between two electrodes.
Second, time series measurement over a single dot
 can be performed.
For this, the procedure of preparing the dot
 to the desired initial state, applying an ESR field and
 measuring the current has to be repeated many times
 (see Sec.~\ref{ssecCounting} for counting statistics
 of the read-out process).
Then, assuming ergodicity, the current average
 of all these individual measurements corresponds
 to the ensemble averaged value.

\subsection{Decoherence in the Time Domain}
\label{ssecTTwoTimeDomain}
In Fig.~\ref{figRabiOsc}, we plot the numerical solution of
 Eqs.\ (\ref{eqnRhoDotuu})--(\ref{eqnRhoDotSd}),
 showing the coherent oscillations of $\rhoSys$ and $I_l$,
 for (a) spin-polarized and (b) unpolarized leads.
The decay of these oscillations is
 dominated by the spin decoherence rate $\gammadu$.
Since this decay can be measured via the current,
 $\gammadu$ (and $1/T_2$) can be accessed directly in the time domain 
 (see also Sec.~\ref{secPulsedESR}, Ref.~\onlinecite{SpinEchoFootnote}
  and Fig.~\ref{figRabiOscPulsed}).

\subsection{Zeno Effect}
\label{ssecZeno}

When the rate for electrons tunneling onto the dot, $\WSsigma$,
 is increased, the coherent oscillations 
 of $\rhouu$, $\rhodd$ become suppressed
 (see inset of Fig.~\ref{figRabiOsc}(a)).
This suppression is caused by the increased spin decoherence 
 rate $\gammadu$ [Eq.~(\ref{eqnGammdu})]
 and can be interpreted as a
 continuous strong measurement of the dot-spin,
 performed by an increased number of charges tunneling onto the dot.
This suppression of coherent oscillations is known as Zeno effect.\cite{Peres}
Since it is visible in $\rhoSys$,
 it can be observed via the currents $I_{1,2}(t)$.

     \section{Pulsed ESR and Rabi Oscillations}
\label{secPulsedESR}

We now show that it is possible to observe the coherent Rabi oscillations
 of a single electron spin even without the requirement of 
 measuring time-resolved currents.
This can be achieved by applying ESR pulses of length $t_p$
 and by measuring time-averaged currents 
 (over arbitrarily long times).
Then, the time-averaged current $\bar{I}(t_p)$ 
 as function of $t_p$ gives access to the 
 time evolution of the spin-state of the dot,
 for both, polarized and unpolarized leads.\cite{NakamuraFootnote}
In particular, 
 since arbitrarily long times,
 and thus a large number of electrons, can be used to
 measure $\bar{I}$,
 the required experimental setups are significantly simpler
 compared to setups which aim at measuring time-dependent currents
 with high resolution.

\begin{figure}[t]
\centerline{\psfig{file=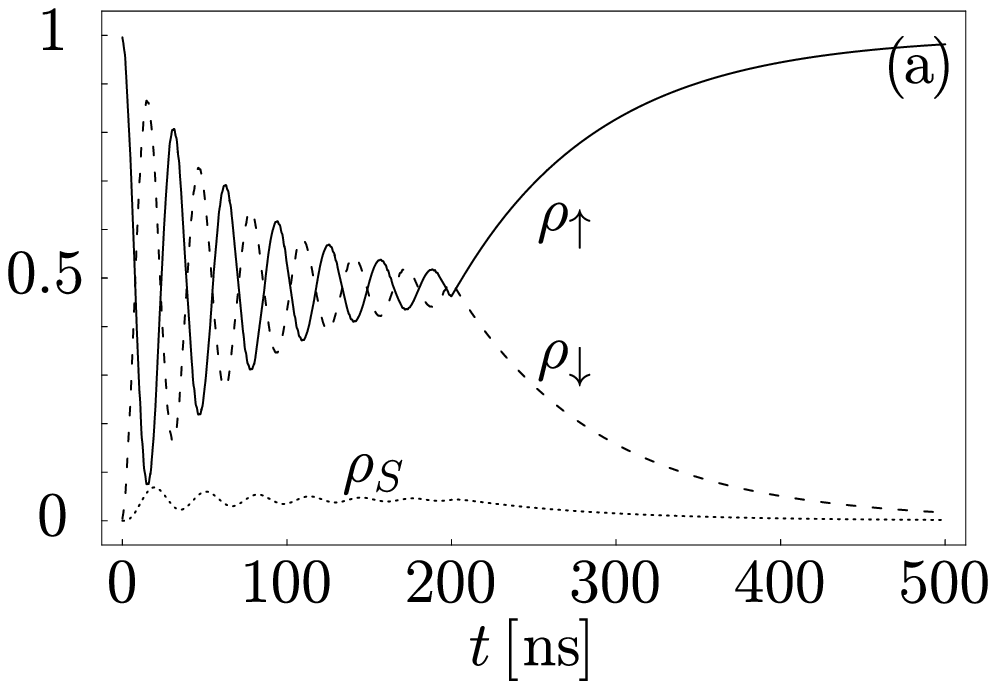,width=\figwidthSmall}}
\centerline{\psfig{file=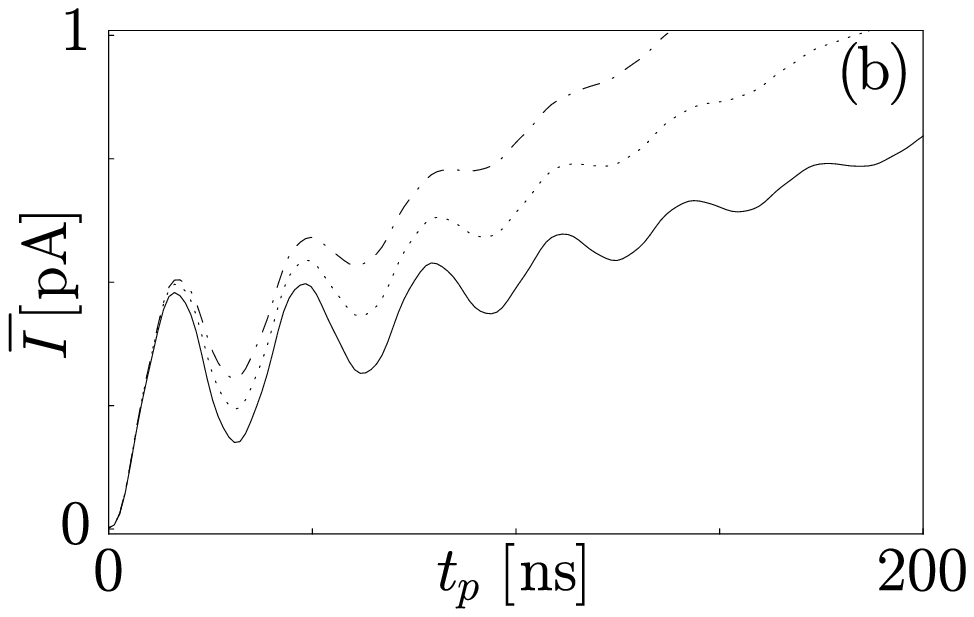,width=\figwidthSmall}}
\centerline{\psfig{file=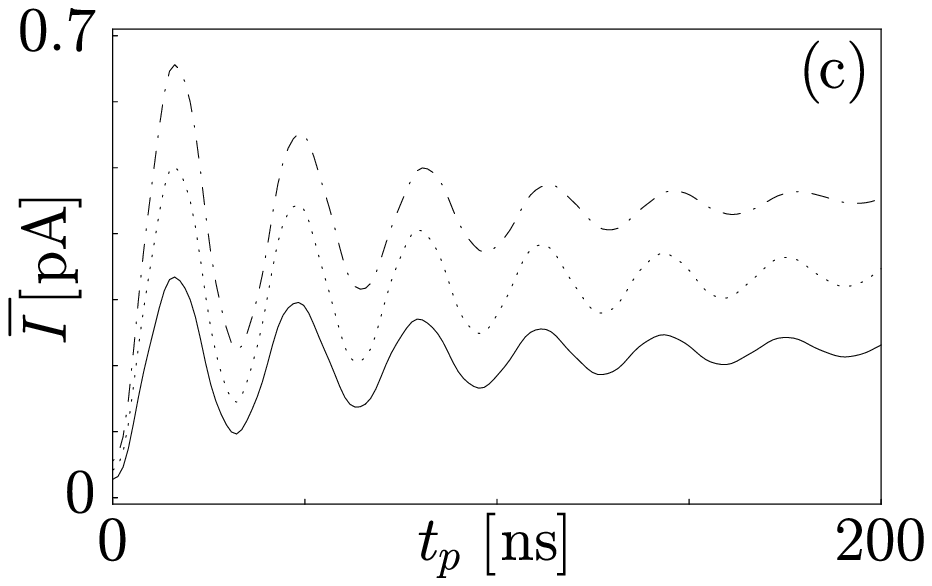,width=\figwidthSmall}}
\caption[Rabi oscillations]{\label{figRabiOscPulsed}
Single spin Rabi oscillations in the current $I(t_p)$ generated by
 ESR pulses of length $t_p$.
Here,
 $\Dm>kT$, 
 Rabi frequency $\HrfConst = 4\times10^8\:{\rm s}^{-1}$
 (corresponding to $g=2$ and $B_x^0=20 \: {\rm G}$),
 $\gO=2\times10^7\:{\rm s}^{-1}$, 
 $\gT=5\gO$,
 $T_1 = 1  \: \mu{\rm s}$, and
 $T_2 = 150\: {\rm ns}$.
(a) Evolution of the density matrix for unpolarized leads
    where a pulse of length $t_p=200\:{\rm ns}$ is switched on at $t=0$,
    obtained by numerical integration of the master equation
    [Eqs.~(\ref{eqnRhoDotuu})--(\ref{eqnRhoDotdu})].
(b) Time-averaged current $\bar{I}(t_p)$ (solid line) for unpolarized leads
  and a pulse repetition time $t_r = 500\: {\rm ns}$.
We also show the current
 where $\gO$ and $\gT$ are increased by a factor of 1.5 (dotted)
 and 2 (dash-dotted).
(c) Time-averaged current $\bar{I}(t_p)$ (solid line) for spin-polarized leads,
 $\guO=2\times10^7\:{\rm s}^{-1}$, $\guT=5\guO$, $\gdl{1,\,2}=0$.
 The pulse repetition time $t_r = 10\:\mu{s}$ is chosen larger than
 $T_1$.
Again, we show the current for tunneling rates $\gul{1,\,2}$
 increased by a factor of 1.5 (dotted) and 2 (dash-dotted).
Note that in this figure $t_p<T_1$, i.e., for most electrons tunneling
 through the dot after the pulse is switched off,
 the linear background is negligibly small.
}
\end{figure}

We assume a rectangular envelope for the ESR pulse 
 with length $t_p$ and repetition time $t_r$ (thus $t_p<t_r$).
The time when no ESR field is present, $t_r-t_p$,
 should be long enough such that the dot can relax into its
 ground state $\spup$, 
 i.e., at the beginning of the next pulse we have $\rhouu=1$.
We calculate $\bar{I}(t_p)$ 
 by numerical integration of the master equation 
 [Eqs. (\ref{eqnRhoDotuu})--(\ref{eqnRhoDotdu})]
 and by subsequently averaging the (time-dependent)
 current [Eq.~(\ref{eqnSTCurrentFromRates})] 
 over the time interval $[0,\, t_r]$.
The results are shown in Fig.~\ref{figRabiOscPulsed}(b) for
 unpolarized leads 
 at the spin satellite peak (see Sec.~\ref{ssecSpinSatellite}),
 and in Fig.~\ref{figRabiOscPulsed}(c) for spin-polarized leads
 in the regime for spin-read out (see Sec.~\ref{secReadOut}).
In both cases, $\bar{I}(t_p)$ as function of pulse length $t_p$
 shows the Rabi oscillations of the dot spin, i.e.,
 the Rabi oscillations can be observed in the time domain even
 without time-resolved measurements.

In addition to the exact numerical evaluation of the master equation
 (see Fig.~\ref{figRabiOscPulsed}),
 we now give an approximate analytical expression for 
 $\bar{I}(t_p)$.
We first consider the case of unpolarized leads at 
 the spin satellite peak (Sec.~\ref{ssecSpinSatellite});
 for the case of spin-polarized leads see below.
For this, we need to evaluate
 the time-average of Eq.~(\ref{eqnSTCurrentFromRates})
 for $kT<\Dm$, 
\begin{equation}
\label{eqnIPulsedAvgT0}
\bar{I}(t_p)= e\,(\guT+\gdT) \:\frac{1}{t_r}\int_0^{t_r}\!dt\: \rhoSS(t) \:.
\end{equation}
First, we consider times $t$ with $0\leq t\leq t_p$, 
 for which an ESR field is present, and $\rhoSys$ oscillates
 with Rabi frequency $\Delta_x$ 
 (see Fig.~\ref{figRabiOscPulsed}(a) for $t\leq 200\:{\rm ns}$).
Qualitatively speaking, 
 when $\rhoSS(t)$ is integrated in Eq.~(\ref{eqnIPulsedAvgT0})
 up to $t_p$,
 the oscillating contribution averages nearly to zero,
 and we obtain a background contribution $\Ibg$
 approximately proportional to $e(\guT+\gdT) t_p/t_r$, i.e.,
 linear in $t_p$,
 in agreement with Fig.~\ref{figRabiOscPulsed}(b).
For experiments,
 this linearity of $\Ibg$ provides a first check
 that $t_r$ is sufficiently long
 such that the dot has indeed relaxed into its ground state
 before the next pulse is applied.
We also give an upper bound for $\Ibg$
 by using the inequality
 $\rhoSS \leq \rhoSSmax = \WSd/(\WSd+\WSu+\WuS)$.
This is seen as follows.
 For $\rhoSS(t)>\rhoSSmax$, we would have $\rhoDotSS(t)<0$, and
 thus $\rhoSS(t')>\rhoSSmax$, for all $0\leq t' \leq t$,
 which would be in contradiction to the initial condition $\rhoSS(0)=0$,
 hence indeed $\rhoSS(t)\leq \rhoSSmax$.
From Eq. (\ref{eqnIPulsedAvgT0}), we then obtain 
 $\Ibg < e \min\{\guO,\, \guT+\gdT\}\, t_p/t_r$.
Note that for pulse lengths $t_p$, over which the dot 
 spin evolves coherently, $t_p \guO \lesssim 1$.
 Thus, by comparing the upper bound with Eq.~(\ref{eqnIAvgAfterPulse}),
 we see that for $\guO<\gT$
 the background current $\Ibg$ never becomes dominant.

Second, we consider $t_p\leq t\leq t_r$, i.e.,
 the ESR field is switched off, and
 the dot state relaxes into its ground state $\spup$.
Making the reasonable assumption that the tunnel processes 
 dominate the spin relaxation, $\gAvg>\Wud$, 
 we neglect $\Wud$ here.
We then calculate the contribution for $t\geq t_p$
 to the integral in Eq.~(\ref{eqnIPulsedAvgT0})
 analytically, and obtain (up to $\Ibg$)
\begin{equation}
\label{eqnIAvgAfterPulse}
 \bar{I}(t_p) \approx \frac{e}{t_r}\,
 \frac{\guT+\gdT}{\gdO+\gdT}\: [\rhodd(t_p)+\rhoSS(t_p)] 
  \:\:\propto\:\: 1-\rhouu(t_p)\,.
\end{equation}

We now give a physical explanation for Eq.~(\ref{eqnIAvgAfterPulse}).
We consider different tunneling events
 (after the pulse is switched off)
 and their contributions to the current,
 $\int_{t_p}^{t_r} dt\: \rhoSS(t)$.
Since we assume that at $t_r$ the dot 
 has relaxed into its ground state $\spup$, and
 thus $\rhoSS(t_r)=\rhodd(t_r)=0$,
 it is sufficient to consider only one pulse and to extend the
 upper integration limit to infinity.
For the population $\rhodd(t_p)$ of state $\spdown$,
 the only allowed transition is $\spdown\to\ket{S}$
 (neglecting again the intrinsic spin relaxation rate $\Wud$).
Thus, eventually this population $\rhodd$ 
 will be transfered to $\rhoSS$ and thus to the current.
Note that sequences with $\ket{S}\to\spdown$ contribute 
 to the current at a later time again, 
 since the only possible decay into the ground state 
 $\spup$ involves $\ket{S}$.
Therefore, concerning current contributions,
 we introduce the effective population $\rho_I=\rhodd+\rhoSS$,
 which is the probability that at some later time an electron
 can still tunnel from the dot to lead 2.
This $\rho_I$ decays to state $\spup$ with the
 rate $\gamma_S = \gamma_1^\downarrow+\gamma_2^\downarrow$,
 i.e., with the rate for the process $\ket{S}\to\spup$.
In total, integrating over $\rhoSS(t)$ for $t>t_p$
 yields 
 $\int_0^\infty\,dt \: \rho_I(t_p)e^{-\gamma_S t}=
  [\rhodd(t_p)+\rhoSS(t_p)]/\gamma_S$,
 and with Eq.~(\ref{eqnIPulsedAvgT0}) we immediately
 recover Eq.~(\ref{eqnIAvgAfterPulse}), as expected.

Next, we consider the case for spin-polarized leads.
Here, no spin relaxation process due to tunneling occurs
 and the dot-spin can only relax via intrinsic spin flips,
 given by the rate $\Wud$
 (corresponding to the relaxation time $T_1$;
 we neglect $\Wdu$ since $\Wdu\ll\Wud$).
Thus, we now consider the relaxation rate $\Wud$ instead of $\gamma_S$.
The relaxation occurs only from $\spdown$ to $\spup$,
 i.e., the roles of $\ket{S}$ and $\spdown$ are interchanged
 compared to the case for unpolarized leads considered above.
The above argument now applies analogously 
 by considering the (spin-polarized) current in lead 1,
 $I_1^\uparrow(t)=e\,\guO\,\rhodd(t)$.
We obtain
\begin{equation}
\label{eqnIAvgAfterPulseSM}
 \bar{I}^\uparrow(t_p) \approx \frac{e}{t_r}\,
 \frac{\guO}{\Wud}\:[1-\rhouu(t_p)],
\end{equation}
 with equality for $t_p\ll T_1$.
We point out that for $\guO\gg 1/T_1$,
 the decoherence of the dot-spin occurs much faster than
 its relaxation.
Then, for pulse lengths $t_p$, for which Rabi oscillations
 can be observed, are limited, $1/t_p \gtrsim\gammadu>\guO\gg\Wud$.
In this case, the current contribution for $t\leq t_p$ can be neglected
 since they are suppressed by a factor of $t_p\Wud\ll1$
 compared to the contribution for $t\geq t_p$
 [Eq.~(\ref{eqnIAvgAfterPulseSM})],
 see Fig.~\ref{figRabiOscPulsed}(c).
Note that for spin-polarized leads,
 the relaxation time, $\Wud^{-1}$,
 is usually much longer than for unpolarized leads, $\gamma_S^{-1}$,
 thus the required pulse repetition time $t_r>\Wud^{-1}$ might become very long.
However, if one chooses a pulse repetition time $t_r=c/\gamma$, for $c>1$, and
 with the relevant relaxation rate $\gamma$, the current is
 proportional to $(1/t_r)\int_0^\infty\!dt\:e^{-\gamma t}=1/c$,
 i.e., independent of $\gamma$.
Thus, roughly speaking, the slow relaxation rate in the case of spin-polarized
 leads has no influence on the attainable maximum
 current since the decay from $\rhoSS$ and $\rhodd$
 is much slower and thus per pulse there are more electrons 
 passing the dot.

{}To conclude, we would like to emphasize again that the Rabi oscillations of
 the dot-spin can be observed directly in the time domain
 by using pulsed ESR and measuring time-averaged currents
 (see Fig.~\ref{figRabiOscPulsed}).
Observing Rabi oscillations also allows to determine $T_2$ in
 the time domain, see Sec.~\ref{ssecTTwoTimeDomain}.\cite{SpinEchoFootnote}

             \section{STM Techniques and ESR}
\label{secSTM}

So far, we have considered a quantum dot coupled to leads.
In this section, we would like to note that our description 
 applies to more general structures
 showing Coulomb blockade behavior,
 such as Au nanoparticles\cite{Schoenenberger}
 or ${\rm C}_{60}$ molecules,\cite{PorathC60}
 which has been observed with STM techniques.
This justifies that instead of a quantum dot,
 we now consider a localized surface state or
 an atom, molecule, or nanoparticle adsorbed on a substrate.
This particle can then be probed
 with the STM tip by
 measuring the tunnel current through the particle.
The current arises from
 electrons tunneling from the STM tip onto the particle, 
 and further tunneling, possibly through an insulating overlayer,
 into the bulk of the substrate.

In standard STM theory, 
 the tunneling from the STM tip to the sample is
 treated pertubatively.\cite{ChenSTM}
Evaluation of the golden rule matrix element,
 in the simplest model of a one-dimensional tunnel barrier,
 gives a tunneling amplitude, which
 is dominated by an exponential decay of
 the electronic wavefunction into the barrier,
 thus $t_l^\sigma\propto e^{-\kappa d}$ [cf. Eq.~(\ref{eqnStRateZeroT})], 
 with $\kappa=\sqrt{2m\phi}$, tip--particle distance $d$, 
 and barrier height $\phi$ 
 (roughly given by the work function of the tip/sample).
In particular, the perturbative description of STM is equivalent to
 our treatment of the tunneling Hamiltonian in first (sequential tunneling)
 order.
Therefore,
 if the particle of interest shows Coulomb blockade behavior and
 has a spin-$\frac{1}{2}$ ground state, the master equation
 [Eqs.~(\ref{eqnRhoDotuu})--(\ref{eqnRhoDotSd})] applies.
Thus, using an ESR field,
 coherent Rabi oscillations and
 the $T_2$ time of the spin state of the particle 
 can be accessed via the current.
Further, if spin-polarized tips and/or substrates are available
 (spin-polarized STM),
 such a particle can act as single spin memory with
 read-out via current.
Note that the tunneling rates from the STM tip into the particle can be 
 controlled by changing the distance $d$, thus
 the total decoherence $\gammadu$  [Eq.~(\ref{eqnGammdu})],
 containing tunneling contributions, can be varied.
This allows, e.g.,
 to vary the current linewidth, $2\gammadu$, (Sec.~\ref{ssecMeasureTtwo}),
 and to suppress the Rabi spin flips for strong decoherence
 (Zeno effect, Sec.~\ref{ssecZeno}).
One apparent restriction of atomic or molecular systems is that 
 it is difficult to apply a gate voltage to the particle,
 shifting its energy levels.
However, the same effect can be achieved if the Fermi
 energies in the STM tip and the substrate can be shifted,
 such as by varying electron densities.

                       \section{Discussions}
\label{secDisc}

We have shown how single spin dynamics
 of quantum dots can be accessed by current measurements.
We have derived and analyzed coupled master equations of a quantum dot,
 which is tunnel coupled to leads, 
 in the presence of an ESR field.
The current through the dot in the sequential tunneling regime shows a 
 new resonance peak (satellite peak)
 whose linewidth provides a lower bound on the single 
 spin decoherence time $T_2$.
We have shown that also the cotunneling current 
 has a resonant current contribution, giving access to $T_2$.
The coherent Rabi oscillations of the dot-spin can be observed by charge
 measurements, since they lead to oscillations 
 in the time-dependent current and 
 in the time-averaged current as function of ESR pulse length.
We have shown how the ESR field can pump current through a dot
 at zero bias if spin dependent tunneling or
 a spin inverter is available.
We have discussed the concept of measuring a single spin via
 charge in detail.
We have identified the measurement time
 of the dot-spin via spin-polarized leads.
Finally, we have noted that the concepts presented here
 are not only valid for quantum dots
 but also for ``real'' atoms or molecules if they
 are contacted with  an STM tip.

                          \begin{acknowledgments}
We thank
 C. Bruder,
 G. Burkard,
 S. De Franceschi, 
 J.M. Elzerman,
 H. Gassmann,
 R. Hanson,
 L. Kouwenhoven,
 M. Leuenberger, 
 F. Marquardt,
 F. Meier,
 S. Oberholzer,
 P. Recher,
 E. Sukhorukov,
 W.G. van der Wiel,
 and M.R. Wegewijs
 for discussions.
We acknowledge support from the Swiss NSF, DARPA, and ARO.
\end{acknowledgments}

\appendix

                       \section{Stationary current}
\label{appStatCurrent}

Here, we give the various formulas for the stationary
 current through the dot 
 in the sequential tunneling regime
 and in the presence of an ESR field.
We have calculated the current by evaluating the stationary
 solution of the master equation (Sec. \ref{secStationaryCurrent}) 
 and with Eq.~(\ref{eqnSTCurrentFromRates}).
For odd-to-even sequential tunneling,
 the spin $\uparrow$ polarized current in lead 2 is

\begin{widetext}

\begin{eqnarray}
I^{\uparrow}_2&=&  \nonumber
e\guT 
\Bigglb(
\guO\sum\limits_{l,l'}
 (-1)^l \gdl{l'} \fdl{l} \ful{l'}
+ \sum\limits_{l}
  \frac{2\Womegarf+\Wud+\Wdu}{2} \left\{
  (-1)^l  \guO \fdl{l}+ \gdl{l}[\fdl{2}-\ful{l}] \right\}
\\ \nonumber &&\qquad
- \sum\limits_{l}\frac{\Wud-\Wdu}{2} \left\{
   (-1)^l \guO \fdl{l}  +
   \gdl{l} [\fdl{2} + \ful{l} - 2\fdl{2}\ful{l}]
  \right\}
\Biggrb)
\\ && \label{eqnFullStatCurrent}
\times \Bigglb(
\sum\limits_{l,\,l'} \gul{l}\gdl{l'} \,\left\{
  1- \left[1-\fdl{l}\right]\left[1-\ful{l'}\right] \right\}
\:+\:
\sum\limits_{l,\, \sigma\neq\sigma'} (\Womegarf+\WP_{\sigma'\sigma})
 \left\{ \gls{l}{\sigma} + \gls{l}{\sigma'}[1-\fls{l}{\sigma}] \right\}
\Biggrb)^{-1}.
\end{eqnarray}
\end{widetext}

\noindent
The spin $\downarrow$ polarized current, $I^{\downarrow}_2$, is obtained 
 from Eq.~(\ref{eqnFullStatCurrent}) by exchanging
 all $\uparrow$ and  $\downarrow$ in the numerator
 (the denominator remains unaffected by such an exchange).
The currents in lead 1, $I_1^{\uparrow,\downarrow}$ are obtained
 from the formulas for $I_2^{\uparrow,\downarrow}$
 by exchanging indices 1 and 2 and by a global change of sign.
The charge current is $I_l=\sum_\sigma I_l^\sigma$
 and is equal in both leads, $I=I_1=I_2$,
 due to charge conservation.
For large Zeeman splitting $\Delta_z>\Delta\mu,\, kT$
 and around the spin satellite peak,  $\mu_1>\ESd>\mu_2$
 (see Sec.~\ref{ssecSpinSatellite}),
 we have $f_l(\ESu)=0$, and the current is
\begin{eqnarray}
\label{eqnCurrentSatelliteSpinDep}
 I&&=  e \left({\Womegarf} + \Wdu\right)  \nonumber
  \Big[
 (\guO\guT+\guO\gdT)\, f_1(\ESd)
\\ \nonumber
 && \quad -(\gdO\guT+\guO\guT)\, f_2(\ESd)
        \Big] 
\\ \nonumber
 && \times
  \Big\{
  \left( 2\gdAvg - \Wud - {\Womegarf} \right) 
  \left[\guO f_1(\ESd) + \guT f_2(\ESd) \right] \,
\\ 
 && \quad +
    2\left( \Wud+\Wdu+2{\Womegarf} \right) 
     \left( \gdAvg +  \guAvg \right)
 \Big\}^{-1}
{}\:,
\end{eqnarray}
for which we have given special cases in 
 Eqs.~(\ref{eqnCurrentSatellite}), (\ref{eqnCurrentZeroT}),
     (\ref{eqnCurrentLinResp}) and (\ref{eqnCurrentDeltaMuZero}).

For completeness, 
 we also give the results for even-to-odd sequential tunneling,
 as discussed in Sec.~\ref{secZeroOne}.
By applying the replacements given in  Sec.~\ref{secZeroOne}.
 to Eq.~(\ref{eqnFullStatCurrent}),
 we obtain the spin $\downarrow$ polarized stationary current in lead 2,

\begin{widetext}

\begin{eqnarray}
 \nonumber
I^{\downarrow}_2&=&
e\gdT 
\Bigglb(
\gdO\sum\limits_{l,l'}
 (-1)^l \gul{l'} \fZdl{l} [1-\fZul{l'}]
+ \sum\limits_{l}
  \frac{2\Womegarf+\Wud+\Wdu}{2} \left\{
  (-1)^l  \gdO \fZdl{l}+ \gul{l}[\fZdl{2}-\fZul{l}] \right\}
\\ \nonumber &&\qquad
- \sum\limits_{l}\frac{\Wud-\Wdu}{2} \left\{
   (-1)^l \gdO \fZdl{l}  -
   \gul{l} [\fZdl{2} + \fZul{l} - 2\fZdl{2}\fZul{l}]
  \right\}
\Biggrb)
\\  &&
\times \Bigglb(
\sum\limits_{l,\,l'} \gdl{l}\gul{l'} \,\left\{
  1- \fZdl{l}\fZul{l'} \right\}
\:+\:
\sum\limits_{l,\, \sigma\neq\sigma'} (\Womegarf+\WP_{\sigma'\sigma})
 \left\{ \gls{l}{\sigma'} + \gls{l}{\sigma}\fZls{l}{\sigma} \right\}
\Biggrb)^{-1}\:.
\label{eqnFullStatCurrentZeroOne}
\end{eqnarray}

\end{widetext}

\clearpage



\begin{thebibliography}{}
\bibitem{Secular}
At this point,
 one usually makes a secular approximation\cite{Blum} by neglecting
 terms in $M^{\rm I}$ [Eq.~(\ref{eqnMelements})]
 that rapidly oscillate with the energy difference of
 different states on the dot.
However, it is noteworthy that
 for the system considered here,
 no such terms appear and thus no such secular approximation is required.
\bibitem{PATFootnote}
PAT effects are due to the electric rf field component,
  see Sec.~\ref{ssecModelH},
  and thus are less sensitive to changes in $B_z$,
  compared to the magnetic  ESR effects described here.
\bibitem{ABFootnote}
It has been suggested to test the functionality of spin filtering by
 an electron interference experiment with 
 an Aharonov-Bohm (AB) ring,
 with one dot in each arm of the ring.\cite{FilterABRing} 
On the one hand, if both dots allow the same spin to pass,
 the flux enclosed by the electron moving through the upper as well 
 as through the lower arm will lead to interference and the resulting
 AB-oscillations are observable in the current.
On the other hand, if the dots filter out different electron spins,
 say, all spin-$\uparrow$ electrons go through the upper arm
 while all spin-$\downarrow$ electrons go through the lower arm of the ring,
 no magnetic flux is enclosed by neither spin current and
 thus no interference occurs.
However, there is still a leakage (cotunneling) current of the opposite spin 
 through the spin filters,\cite{Recher}
 thus there is still a small current contribution 
 showing AB-oscillations.
\bibitem{FootnoteEigenvects}
Generically this is satisfied.
However, if there is no complete basis of eigenvectors of ${\cal M}$,
 there is power law admixture.
For instance, for $\HrfRotConst=\Sigma_W-\gammadu$, the last
 two eigenvalues, $\lambda_\pm$, in Eq. (\ref{eqnEigenvalues}) are degenerate.
Then, the time evolution of $\brhoSys$
 can contain terms proportional to $t \,e^{\lambda_\pm t}$.
\bibitem{ReadOutFootnote}
Note that by using transport involving many states of the dot,
 the read-out current could be increased and 
 thus the read-out time further decreased.
Further, if instead resonant tunneling
 is considered,
 e.g., into an SET device, faster read-out could be achieved.
\bibitem{FootnoteChargeAccum}
Note that in general $I_1(t) \neq I_2(t)$,
 since charge can accumulate on the dot for some typical time,
 which is limited from above by the inverse tunneling rates.
 Here, charge accumulation occurs with the period 
 of the Rabi oscillations $\HrfConst^{-1}$.
\bibitem{SpinEchoFootnote}
An alternative application of pulsed ESR for
 measuring $T_2$ is spin echo,
 where a well-defined sequence of pulses is applied to cancel the  
 effect of dephasing.\cite{SpinEcho,T2Si,Abragam}
At the end of the sequence,
 the dot spin (in the $\sigma_z$ basis)
 can be measured by one of the techniques discussed in
 Secs.~\ref{secReadOut}, \ref{lblEnsembleAvg}, and \ref{secPulsedESR}.
Here, spin echo compensates for dephasing
 of an array of quantum dots in an inhomogeneous magnetic field $B_z$,
 or in time series measurements (on a single dot),
 where the magnetic field $B_z$ fluctuates in time.
More care must be taken for inhomogeneities in the $g$ factor (Rabi frequency $\HrfConst$), 
 since then a pulse does not rotate the spin by an equal
 angle for every dot.
{}To compensate this, one could, e.g., use
 Carr-Purcell-Meiboom-Gill pulse sequences.\cite{SpinEcho}
\bibitem{NakamuraFootnote}
We note
 that similar concepts
 have been used in recent experiments,\cite{Nakamura}
 however, with a separation of timescales,
 $T_\Omega, t_p \ll \gAvg^{-1}$.
We show here
 that such concepts can be applied more generally 
 for regimes where $T_\Omega, t_p \lesssim \gAvg^{-1}$.
\end{thebibliography}
\end{document}